\documentclass[aps,pre,twocolumn,groupedaddress,showpacs]{revtex4}
\usepackage{graphicx}

\begin{document}


\title{Surface critical behavior of fluids: Lennard-Jones fluid near weakly attractive substrate}



\author{I.Brovchenko}
\email{brov@heineken.chemie.uni-dortmund.de}
\author{A.Geiger}
\email{alfons.geiger@udo.edu}
\affiliation{Physical Chemistry, Dortmund University, 44221 Dortmund, Germany}
\author{A.Oleinikova}
\email{alla.oleinikova@ruhr-uni-bochum.de}
\affiliation{Physical Chemistry, Ruhr-University
Bochum, 44780 Bochum, Germany}

\date{\today}

\begin{abstract}
The phase behavior of fluids near weakly attractive substrates is studied by computer simulations of the coexistence curve of a Lennard-Jones (LJ) fluid confined in a slitlike pore. The temperature dependence of the density profiles of the LJ fluid is found to be very similar to the behavior of water near hydrophobic surfaces (Brovchenko $\textit{et al.}$ J.Phys.: Cond.Matt. $\textbf{16}$, 2004). A universal critical behavior of the local order parameter, defined as the difference between the local densities of the coexisting liquid and vapor phases at some distance z from the pore walls, $\Delta\rho$(z) = ($\rho_l$(z) - $\rho_v$(z))/2, is observed in a wide temperature range and found to be consistent with the surface critical behavior of the Ising model. Near the surface the dependence of the order parameter on the reduced temperature $\tau$ = (T$_c$ - T)/T$_c$ obeys a scaling law $\sim$ $\tau^{\beta_1}$ with a critical exponent $\beta_1$ of about 0.8, corresponding to the $\textit{ordinary}$ surface transition. A crossover from bulk-like to surface-like critical behavior with increasing temperature occurs, when the correlation length is about half the distance to the surface. Relations between the $\textit{ordinary}$ and $\textit{normal}$ transitions in Ising systems and the surface critical behavior of fluids are discussed.
\end{abstract}

\pacs{05.70.Jk, 05.70.Np, 64.60.Fr, 64.70.Dv}

\maketitle

\section{Introduction}
  The presence of a solid boundary effects the properties of a fluid, which becomes spatially heterogeneous normal to the boundary. Besides density oscillations in the close vicinity of the surface due to packing effects, the surface produces perturbations which intrude into the bulk fluid on a distance scale proportional to the bulk correlation length and, therefore, gain increasing importance near the critical point even far away from the surface. The critical behavior of the local properties of the fluid near the surface should follow universal scaling laws, which are different, however, from the scaling laws of the bulk fluid. On strict theoretical grounds simple scaling equations are valid only in the immediate neighborhood of the critical point. However, both experimental and computer simulation studies of the coexistence curves of bulk fluids evidence that contributions from the asymptotic power laws remain dominant over a surprisingly wide temperature range. The same situation could also be expected for the surface critical behavior. In this case the knowledge of the surface critical behavior of fluids opens the possibility to predict the density profiles and related properties of various fluids near solid boundaries in a wide range of thermodynamic conditions. 
\par
Since bulk fluids belong to the universality class of the 3D Ising model, it is natural to map the surface critical behavior of fluids onto the surface universality classes of the Ising model. In the presence of a non-zero surface field (h$_1\neq$ 0, the case relevant for fluids)  in the Ising model a wetting transition ultimately occurs at some temperature below the bulk critical temperature T$_c$ \cite{BinLan1,BinLan2,BinLan3,NakFish}. Above the temperature of the wetting transition the magnetization of the single phase, remaining near the surface, is predicted to follow the law of the so-called $\textit{normal}$ transition \cite{Diehlnorm1,Diehlnorm2} with dominant regular behavior, i.e. with a nonzero constant followed by a leading linear contribution, proportional to the reduced temperature $\tau$ = (T$_c$ - T)/T$_c$, which is a measure of the temperature deviation from the critical temperature T$_{c}$. The available computer simulations of Ising systems at h$_1\neq$ 0 \cite{BinLan1,BinLan2,BinLan3} did not allow to analyze quantitatively the temperature dependence of the surface magnetization both below and above the wetting temperature. 
\par There are no experimental studies of the temperature evolution of the density profile near a solid surface along a liquid-vapor coexistence curve, which would allow to explore the surface critical behavior of one-component fluids. Experimental studies of the order parameter in binary liquid mixtures near a solid surface (difference between the concentrations of the two coexisting phases) evidence that its temperature dependence near the wall follows a power law  $\sim$ $\tau^{\beta_1}$ with a surface critical exponent $\beta_1\approx$ 0.8 \cite{Fenzl1,Fenzl2,Franck}. This value is close to the critical exponent of the $\textit{ordinary}$ transition of the Ising model at h$_1$ = 0, that describes the temperature evolution of the magnetization in a surface layer \cite{BinderHoh1,Binderhoh2}. 
\par
It is usually supposed that the order parameter of one-component fluids and binary fluid mixtures near a surface obeys the laws of the ordinary transition below the wetting (drying) temperature \cite{Pandit,Dietrev,Indekeu}. In the case of a weakly attractive surface the long-range fluid-wall interaction suppresses a drying transition \cite{Indekeu,Ebner1,Ebner2}, and so the order parameter near the surface should follow a temperature dependence  $\sim$ $\tau^{\beta_1}$ up to the bulk critical temperature. 
\par In fact, such a behavior of the order parameter was observed in computer simulation studies of the liquid-vapor phase transition of water near hydrophobic surfaces \cite{PCCP,BGO2004a}. 
The temperature evolution of the densities $\rho_{l,v}$(z,$\tau$) in coexisting liquid (subscript $\textit{l}$) and vapor (subscript $\textit{v}$) phases near a substrate could be described by the following equation:
\begin{eqnarray}
\rho_{l,v}(z,\tau) = (A_0 + A_1 \tau + A_2\tau^2... )\pm B_1 \tau^{\beta_1}, 
\end{eqnarray}
where A$_0$ = $\rho_c(z)$ = $\rho$(z, $\tau$ = 0) is the local density at the critical point, A$_1$, A$_2$... and B$_1$ are local system dependent amplitudes. The asymmetric contribution in the bracket (diameter of the coexistence curve) is the same in both coexisting phases and may include also a singular contribution $\sim$ $\tau$ $^{(2 - \alpha)}$ \cite{Diehlnorm1,Diehlnorm2}, with critical exponent $\alpha$ = 0.109 \cite{beta}. Additionally, a singular contribution due to pressure mixing could also be expected, as in a bulk fluid, where it is $\sim$ $\tau$ $^{2\beta}$ \cite{FisherOrk}. Note, however, that contrary to the bulk case, the exponents of both of these singular terms exceed 1 and, therefore, the regular linear term remains the most important asymptotically. The last term in Eq.(1) represents a symmetric contribution to the densities of the coexisting phases (has opposite signs in two phases), and thus describes the order parameter $\Delta\rho$ = ($\rho_l$ - $\rho_v$)/2, which becomes zero at the critical point at any distance from the surface. The value of the critical exponent $\beta_1$ was found close to the value $\approx$ 0.8 of the $\textit{ordinary}$ transition \cite{PCCP,BGO2004a}.
 \par
 The densities of the coexisting liquid and vapor phases of water in a layer of molecular width near a hydrophobic surface follow Eq.(1) in a wide temperature range: 0.5 $<$ $\tau$ $<$ 0.05 \cite{BGO2004a}. Moreover, the intrusion of the surface critical behavior described by Eq.(1) deeper into the bulk fluid was found to be governed by the bulk correlation length. So far, this was the only computer simulation study of the surface critical behavior of the order parameter of a fluid.
Naturally, the universality of the surface critical behavior, reported for water near a hydrophobic surface \cite{BGO2004a}, should be tested for other fluids and, also, for various strengths of fluid-wall interaction, below as well as above the wetting (drying) temperature. 
\par
In this paper we present a study of the surface critical behavior of a Lennard-Jones (LJ) fluids near a weakly attractive wall by computer simulations of its liquid-vapor coexistence curves in a slitlike geometry. The surface critical behavior is explored by considering the temperature evolution of the density profiles of the coexisting liquid and vapor phases. 
\section{Method}
Gibbs ensemble Monte Carlo (GEMC) simulations \cite{GE} were used to simulate liquid-vapor coexistence curves of bulk and confined LJ fluid having interparticle interactions of the form:
\begin{eqnarray}
U_{LJ}(r) = 4\epsilon \left[\left(\sigma/r\right)^{12}-\left(\sigma/r\right)^6\right],   
\end{eqnarray}
where $\epsilon$ measures the well depth of the potential, while $\sigma$ sets the length scale. The potential was spherically truncated at a radius 2.5 $\sigma$ and left unshifted. No long-range corrections were applied to account for effects of the truncation. The density $\rho$ used in the present paper is the number density scaled by $\sigma^3$, while T is the temperature scaled by $\epsilon$/$\textit{k}_{B}$, where $\textit{k}_{B}$ is Boltzmann`s constant. 
\par Simulations of the bulk liquid-vapor coexistence were performed at 51 temperatures from T = 0.60 to T = 1.17. The two lowest temperatures (T = 0.60 and 0.65) were below the bulk triple-point temperature of the LJ fluid (the values 0.687 \cite{freez1},  0.689 \cite{freez2} and 0.692 \cite{freez3} were reported in the literature). The total number of molecules in liquid and vapor phases N$_l$ + N$_v$ was about 2000 at low and 1500 at high temperatures. The edge size $\textit{L}$ of the cubic simulation box  for the bulk liquid was about 12 $\sigma$ and the number of molecules varied from $\sim$ 1900 to $\sim$ 750 with increasing temperature. The number of molecules in the vapor phase N$_v$ varied from $\sim$ 100 at the lowest studied temperature to $\sim$ 750 near the critical temperature. As a strong difference between the number of particles in the two simulated phases may distort the shape of the coexistence curve \cite{Vall,Vall1}, at high temperatures we always tried to keep N$_v$ close to N$_l$. 
\begin{figure}
\includegraphics[width=8cm]{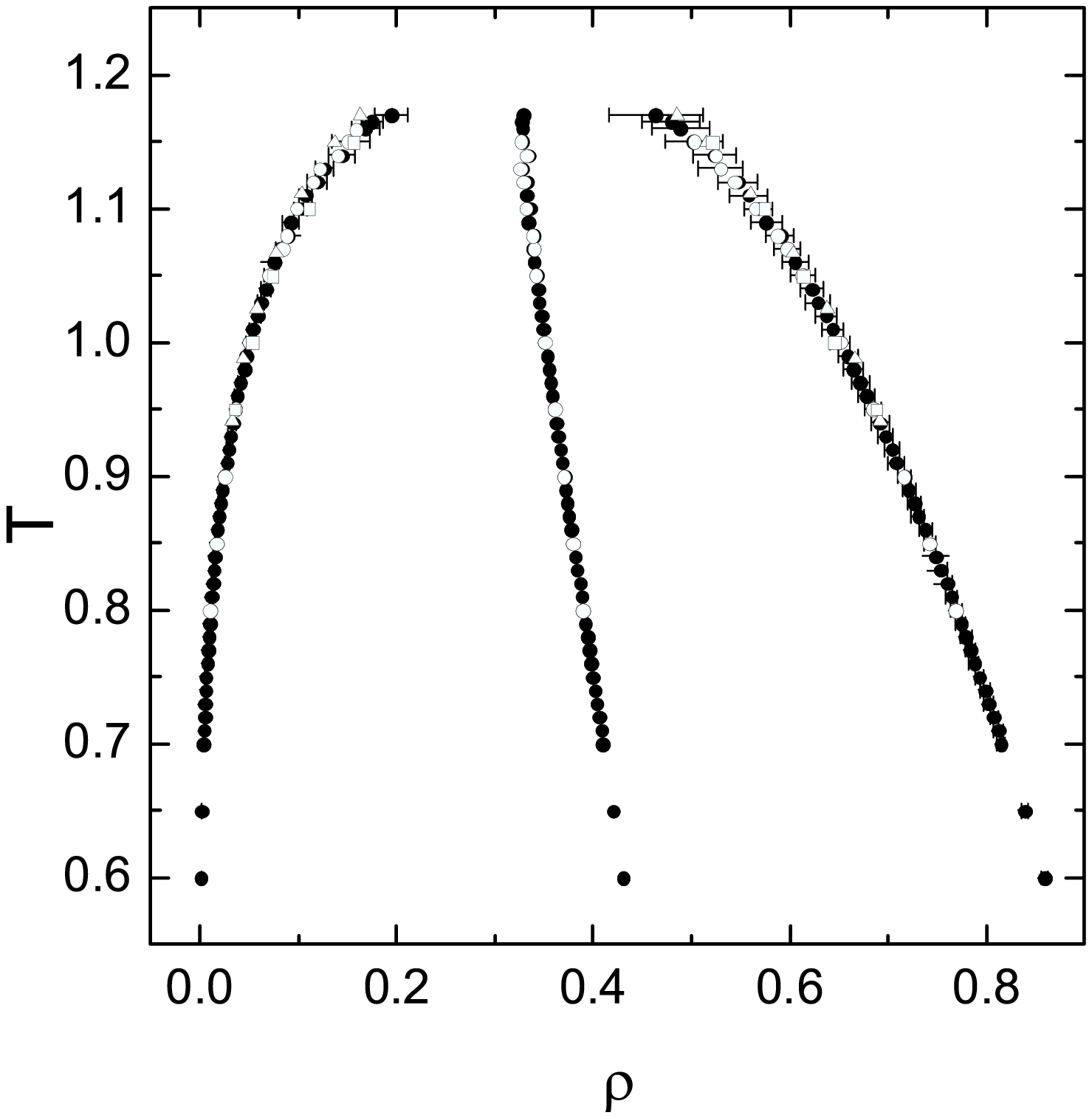}
\caption{Liquid-vapor coexistence curve and its diameter, obtained for the bulk LJ fluid (interaction potential truncated at 2.5 $\sigma$ and unshifted). Closed circles: box size $\textit{L}$ $\approx$ 12 $\sigma$, open circles: $\textit{L}$ $\approx$ 20 $\sigma$. Open squares, triangles and diamonds indicate the data points, reported in Refs.\cite{PanInt94}, \cite{Wilding95} and \cite{Trokhym99}.}
\end{figure} 
\par To estimate the effect of the system size on the coexisting densities, at some temperatures the liquid-vapor coexistence was simulated for larger systems, where the total number of particles in the liquid and vapor phases N$_l$ + N$_v$ was about 7500, which results in a box size $\textit{L}$ for the liquid of about 20 $\sigma$. To test the effect of differing numbers of molecules in the two simulation boxes of the coexisting phases \cite{Vall,Vall1} we repeated the GEMC simulations at T = 1.14 and T = 1.15 with N$_l$ $\approx$ 3N$_v$ and N$_l$ $\approx$ 0.7 N$_v$, respectively. We found this effect negligible, obviously due to the large system size. 
 \par To study the surface critical behavior, the LJ fluid was confined in a slitlike pore with structureless walls. Each wall interacts with particles of the fluid via a long-range potential comprising a single plane of LJ molecules:
\begin{eqnarray}
U_w(z) = 4\epsilon\textsl{ f }\left[0.4 \left(\sigma/z\right)^{10}-\left(\sigma/z\right)^4\right],   
\end{eqnarray}
where z measures the distance to the wall and the parameter $\textsl{f}$ determines the strength of the fluid-wall interaction relatively to fluid-fluid interaction. No truncation was applied to U$_w$(z). In the present paper we report the results obtained for a pore of width $\textit{H}$ = 12 $\sigma$ and $\textit{f}$ = 0.3, which corresponds to a weakly attractive (strongly solvophobic) surface. Note, that this system is very similar to the one, recently studied in equilibrium with the critical bulk LJ fluid \cite{Evans}.
The total number of molecules in the liquid and vapor phases N$_l$ + N$_v$ in the pore varied from 2700 molecules at low temperature to 2000 at high temperatures. The lateral size $\textit{L}$ of the simulation cell for the liquid was about 17 $\sigma$ and so the ratio $\textit{L/H}$ exceeded 1. To explore the effect of the system size, several temperature points were simulated also for a larger system with N$_l$ + N$_v$ of about 8000, that corresponded to $\textit{L}$ $\sim$ 34 $\sigma$.
\par The efficiency of the molecular transfers between the simulation cells was improved by early rejection of insertion attempts which would lead to strong repulsion \cite{BGO2004}. For each temperature point the number of successful transfers per particle between the coexisting phases of the bulk as well as of the confined fluid was about 100 to 200, while the probability of successful transfers varied from 0.5 to 10$\%$. Close to the critical temperature GEMC simulations are limited by identity exchanges between the simulation boxes. This strongly increases the error bars of the estimated densities of the coexisting phases and determines the high-temperature limit of the simulated liquid-vapor coexistence curve. We do not give the densities of the coexisting phases, when identity exchanges occured (except for the highest-temperature point T = 1.17 in the bulk LJ fluid with L $\approx$ 12 $\sigma$). The density profiles of the coexisting liquid and vapor phases and of some supercritical states in the pore were obtained by subsequent Monte Carlo simulations in the NVT ensemble. 
\begin{table}
\caption{Densities of the coexisting phases of the bulk LJ fluid (interaction potential truncated at 2.5 $\sigma$, unshifted), obtained for the system with $\textit{L}$ $\approx$ 12 $\sigma$.}
\label{tab:1}
\begin{tabular}{ccc|ccc} 
\hline\noalign{\smallskip}
T & $\rho_l$ & $\rho_v$ & T & $\rho_l$ & $\rho_v$ \\
\noalign{\smallskip}\hline\noalign{\smallskip}

0.60	& 0.8594(31)	& 97E-5(15)	& 0.94	& 0.6917(90)	& 0.0338(23)\\
0.65	& 0.8386(34)	& 0.00201(17)	& 0.95	& 0.6844(86)	& 0.0359(20)\\
0.70	& 0.8145(41)	& 0.00367(23)	& 0.96	& 0.6783(86)	& 0.0379(22)\\
0.71 &	0.8121(43)	& 0.00434(19)	& 0.97	& 0.6716(87)	& 0.0412(24)\\
0.72	& 0.8075(41)	& 0.00505(20)	& 0.98	& 0.665(10)	& 0.0452(32)\\
0.73 &	0.8017(44)	& 0.00521(43)	& 0.99	& 0.659(10)	& 0.0472(26)\\
0.74 &	0.7984(44)	& 0.00587(33)	& 1.00	& 0.651(10)	& 0.0494(25)\\
0.75 &	0.7927(46)	& 0.00637(65)	& 1.01	& 0.644(11)	& 0.0536(35)\\
0.76 &	0.7876(50)	& 0.00748(55)	& 1.02	& 0.637(11)	& 0.0586(38)\\
0.77 &	0.7833(47)	& 0.00806(38)	& 1.03	& 0.628(13)	& 0.0623(34)\\
0.78 &	0.7785(58)	& 0.00932(86)	& 1.04	& 0.622(12)	& 0.0669(53)\\
0.79 &	0.7741(49)	& 0.0098(11)	& 1.05	& 0.613(12)	& 0.0710(59)\\
0.80 &  0.7686(56)	& 0.0106(12)	& 1.06	& 0.605(13)	& 0.0751(59)\\
0.81 &	0.7642(58)	& 0.0120(11)	& 1.07	& 0.597(13)	& 0.0816(52)\\
0.82 &	0.7597(60)	& 0.0134(12)	& 1.08	& 0.590(14)	& 0.0896(64)\\
0.83 &	0.7533(61)	& 0.0141(14)	& 1.09	& 0.575(16)	& 0.0923(82)\\
0.84 &	0.7480(62)	& 0.0151(14)	& 1.10	& 0.568(15)	& 0.1027(66)\\
0.85 &	0.7425(64)	& 0.0165(15)	& 1.11	& 0.558(19)	& 0.1071(71)\\
0.86 &	0.7379(65)	& 0.0178(12)	& 1.12	& 0.547(20)	& 0.119(10)\\
0.87 &	0.7306(67)	& 0.0194(14)	& 1.13	& 0.529(22)	& 0.127(10)\\
0.88 &	0.7263(68)	& 0.0209(19)	& 1.14	& 0.524(22)	& 0.144(13)\\
0.89 &	0.7208(68)	& 0.0227(18)	& 1.15	& 0.502(30)	& 0.153(19)\\
0.90 & 0.7167(71)	& 0.0244(15)	& 1.16	& 0.489(29)	& 0.168(14)\\
0.91 &	0.7084(82)	& 0.0276(20)	& 1.165	& 0.479(29)	& 0.175(11)\\
0.92 &	0.7038(79)	& 0.0290(17)	& 1.17	& 0.463(48)	& 0.195(17)\\
0.93 &	0.6973(77)	& 0.0307(22)\\					
\noalign{\smallskip}\hline
\end{tabular}
\end{table}
\section{Results}
\subsubsection{Bulk coexistence curve}
The coexistence curve of the bulk LJ fluid was studied by various simulation methods and its critical parameters were found to be strongly sensitive to the details of the interaction potential (cut-off, long-range corrections, use of shifted potential)\cite{Hansen,GE,Lotfi,Smit,Gub93,PanInt94,Debe94,Wilding95,Camp96,Caillol,Trokhym99,Pablo99,Potoff00,Shi,Pablo02,Pan02,Okumura,Er03}. For example, the values of the critical temperature T$_{c}$, reported for a LJ potential, which was truncated at 2.5 $\sigma$ but not shifted (1.1876(3) \cite{Wilding95}, 1.186(2) \cite{Potoff98}, 1.186 \cite{Trokhym99} 1.1879(4) \cite{Shi}), are essentially lower than the values, reported for a full LJ potential (1.310 \cite{Lotfi}, 1.3120(7)\cite{Potoff98}, 1.3145(2) \cite{Shi}), 1.3207(4) \cite{Okumura}). The critical density $\rho_{c}$ is less sensitive to the truncation of the pair potential: values of $\rho_{c}$ from 0.314 to 0.316 were reported for the full potential \cite{Lotfi,Potoff98,Shi,Okumura}, whereas for a LJ potential , which was truncated at 2.5 $\sigma$ and not shifted it varies from 0.316 to 0.320 \cite{Wilding95,Potoff98,Trokhym99,Shi}. 
\par The densities of the coexisting phases of the bulk LJ fluid with truncated (at 2.5 $\sigma$) and unshifted intermolecular potential, obtained in the present studies for the system with L $\approx$ 12 $\sigma$ and L $\approx$ 20 $\sigma$, are plotted in Fig.1. In Table I the values are given for systems with L $\approx$ 12 $\sigma$. Our results are in good agreement with the available data for the same interaction \cite{PanInt94,Wilding95,Trokhym99}, which are also shown in Fig.1.  A more sensitive comparison is based on the dependence of the order parameter $\Delta \rho$ = ($\rho_{l}$ - $\rho_{v}$)/2 on the reduced temperature $\tau$. This needs knowledge of the critical temperature T$_{c}$. The value of T$_{c}$ could be estimated from fits of the order parameter to the extended scaling equation
\begin{eqnarray}
\Delta \rho = B_0 \tau^{\beta}(1+B_\Delta \tau^{\Delta}+B_{2\Delta}
 \tau^{2\Delta} + ...),
\end{eqnarray}
where $\beta$ = 0.326 is the critical exponent of the bulk coexistence curve, $\Delta$ = 0.52  \cite{beta} is a correction-to-scaling exponent, which accounts to the deviations from the asymptotic critical behavior, and B$_i$ are the system-dependent amplitudes. Fits of Eq.(4) with the leading term only (fits 1 and 2, Table II) provide a rather satisfactory description of the data. In particular, if the temperature is fixed at the most accurately estimated value T$_{c}$ = 1.1876 \cite{Wilding95}, the obtained value of $\beta$ = 0.324 is close to the value of the 3D exponent of the Ising system $\beta$ = 0.326. Two correction terms should be included into Eq.(4) in order to achieve the best fits of our simulations data (fits 3 an 4, Table II). "Best" fit means, that addition of further correction terms does not improve its accuracy.
\begin{figure} [htb]
\includegraphics[width=8cm]{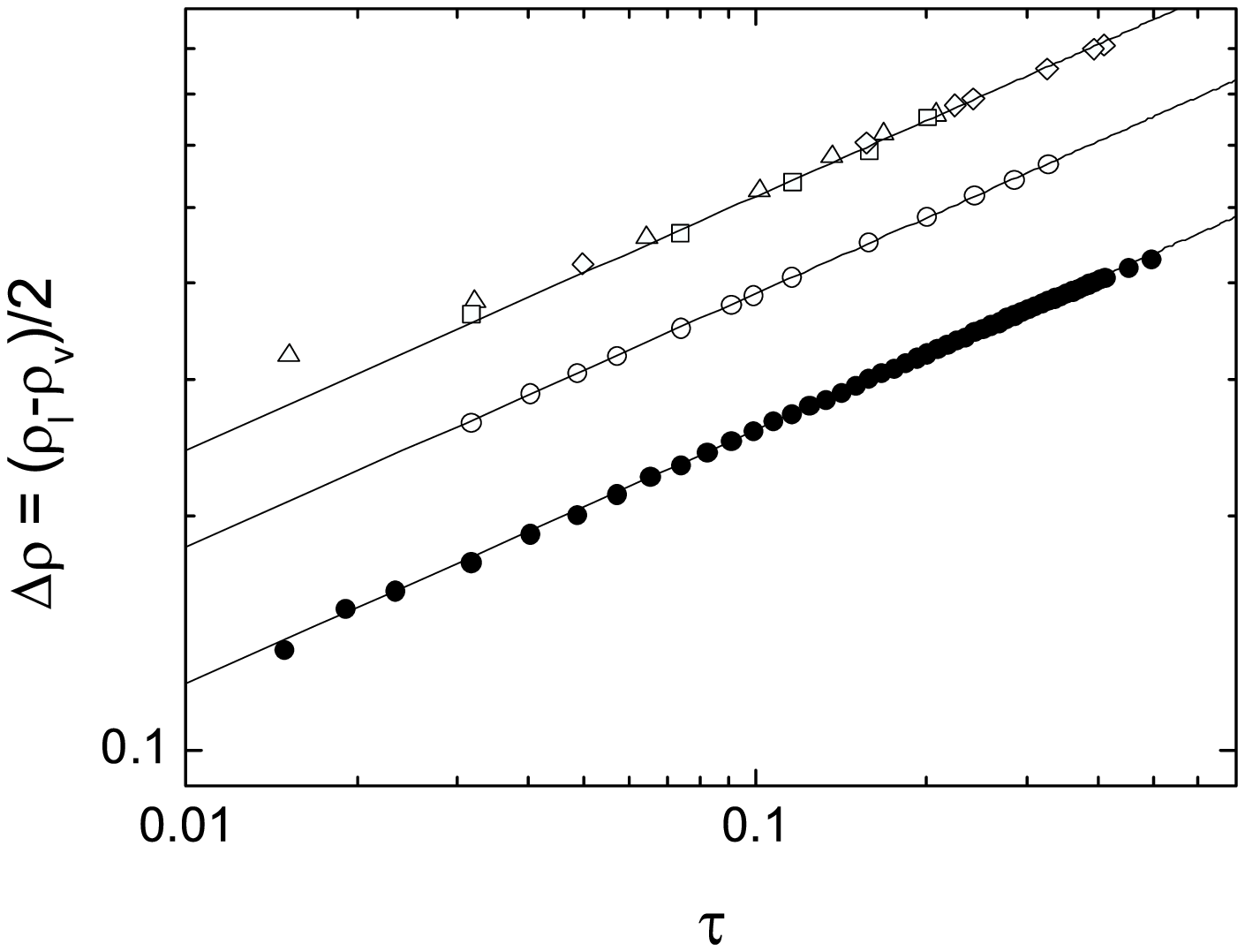}
\caption{Log-log plot of the order parameter $\Delta\rho$ vs reduced temperature $\tau$ for the bulk LJ fluid with L $\approx$ 12 $\sigma$ (solid circles) and L $\approx$ 20 $\sigma$ (open circles, shifted vertically). The data from \cite{PanInt94}, \cite{Wilding95} and \cite{Trokhym99} are also shifted vertically and shown by open squares, triangles and diamonds, respectively. The solid lines show the asymptotic critical behavior of the Ising model with a slope $\beta$ = 0.326 and parameters of fit 2 (Table II).}
\end{figure}
\begin{table}
\caption{Values of the parameters obtained from fits of Eq.(4) to the order parameter $\Delta\rho$. $\chi^2$ represents the mean-square deviations normalized to a dispersion of the simulated data of 0.001 with a confidence limit of 0.95. The fixed parameter are shown italic.}
\label{tab:2}       
\begin{tabular}{ccccccc}
\hline\noalign{\smallskip}
 & $\beta$ & T$_c$ & B$_0$  & B$_\Delta$  & B$_{2\Delta}$ & $\chi^{2}$\\ 
\noalign{\smallskip}\hline\noalign{\smallskip}
\multicolumn{7}{c}{bulk LJ fluid, $\textit{L}$ $\approx$ 12 $\sigma$ } \\
1  &  0.324(1) & $\textit{1.1876}$ & 0.5450(10) &  &  & 2.55 \\
2  &  $\textit{0.326}$ & 1.1873(4) & 0.5460(5) &  &  & 2.60 \\
3  & $\textit{0.326}$ & 1.1881(6) & 0.5247(50) & 0.216(42) & -0.247(40) & 0.74 \\
4  & $\textit{0.326}$ & $\textit{1.1876}$ & 0.5298(21) & 0.179(20) & -0.213(22) & 0.74 \\
\noalign{\smallskip}\hline\noalign{\smallskip}
\multicolumn{7}{c}{bulk LJ fluid, $\textit{L}$ $\approx$ 20 $\sigma$ } \\
5  & $\textit{(0.326)}$ & 0.1892(20) & 0.524(19) & 0.209(70) & -0.233(82) & 4.55 \\
\noalign{\smallskip}\hline\noalign{\smallskip}
\multicolumn{7}{c}{confined LJ fluid, H = 12 $\sigma$, $\textit{L}$ $\approx$ 17 $\sigma$} \\
6 & $\textit{(0.326)}$ & 1.1475(16) & 0.322(13) & 1.30 (21) & -0.895(181)& 2.04 \\
7 & $\textit{(0.125)}$ & 1.1443(25) &0.097(9)& 6.8(1.1) & -3.05(65)  & 2.03 \\
\noalign{\smallskip}\hline
\end{tabular}
\end{table}
The values of the critical temperature T$_{c}$ obtained from the fits of our data (T$_{c}$ = 1.1881(6) for the system with L $\approx$ 12 $\sigma$ and T$_{c}$ = 1.1892(20) for the system with L $\approx$ 20 $\sigma$) are in good agreement with the value T$_{c}$ = 1.1876(3) obtained using a histogram reweighting method with subsequent mixed-field finite-size scaling \cite{Wilding95}.
Use of the latter value of T$_{c}$ does not worsen the quality of the fit (see fit 4 in Table II). Taking into account, that the method used in \cite{Wilding95} provides the most reliable estimates of the critical parameters for the considered LJ fluid, we accept the value of the critical temperature T$_{c}$ = 1.1876 for the subsequent analysis. Note, that a recent improvement of the finite-size scaling method allows a more precise determination of the critical density but does not influence noticeably the critical temperature \cite{Fishermethod}. 
\begin{figure}[htb]
\includegraphics[width=8cm]{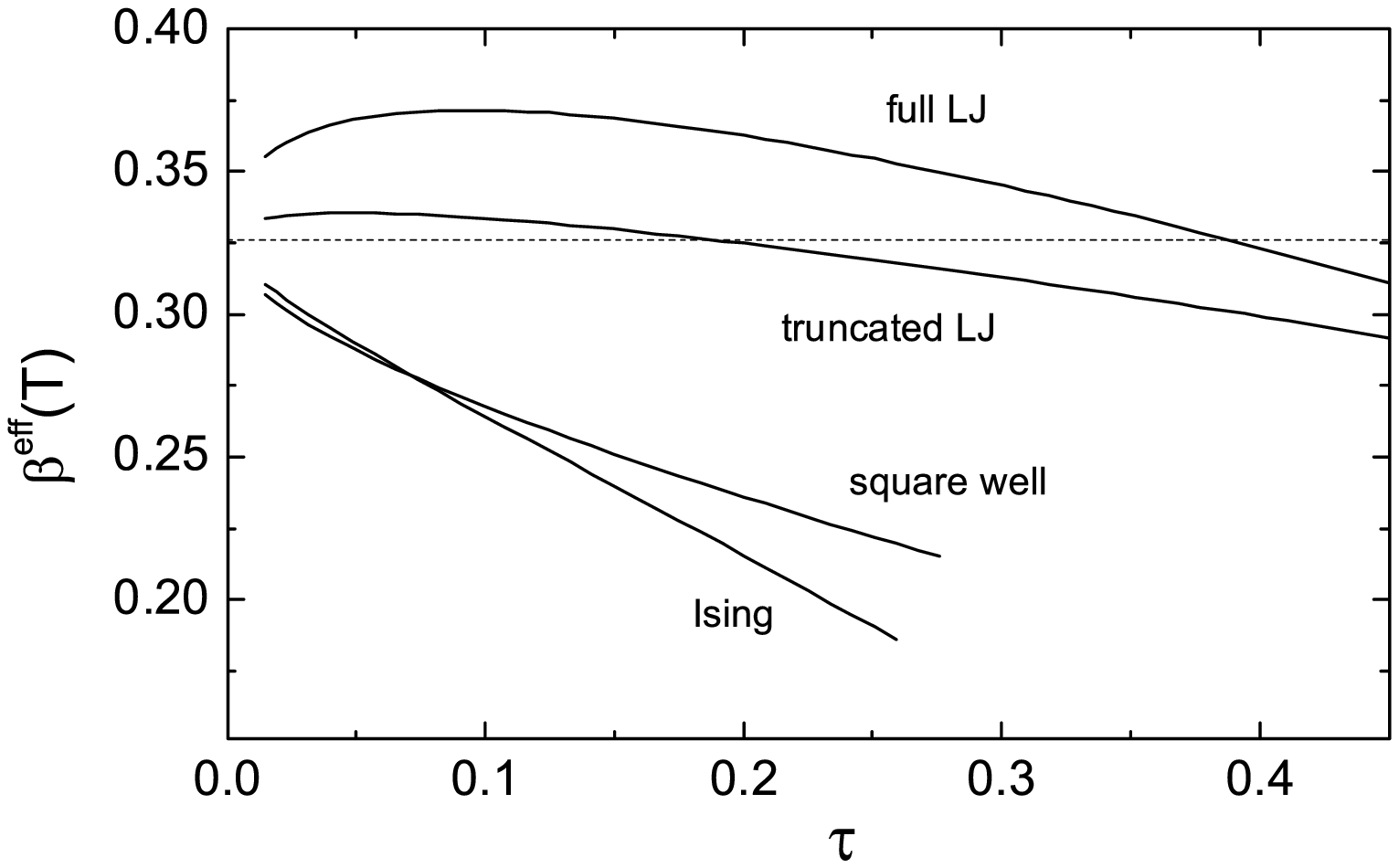}
\caption{The effective critical exponent $\beta^{eff}(\tau)$ defined by Eq.(5) vs reduced temperature $\tau$ for several model systems: Ising model \cite{Isingbeta}, full LJ potential \cite{Lotfi} with critical temperature $T_c$ = 1.312 from \cite{Potoff98}, square-well fluid \cite{FisherSW} and LJ fluid with the potential truncated at 2.5 $\sigma$ (this work).}
\end{figure}
\par  
The temperature dependence of the order parameter $\Delta \rho$ is shown in Fig.2 in double logarithmic scale. All data points from Ref.\cite{PanInt94} and Ref.\cite{Trokhym99} and most of the data points from Ref.\cite{Wilding95} perfectly agree with our data. Two high-temperature points, reported in \cite{Wilding95}, slightly deviate upwards from the asymptotic critical behavior, probably due to the relatively small system size, used in \cite{Wilding95}. Note also, that a slight deviation downward of the highest temperature point in our set for the system with L $\approx$ 12 $\sigma$ may be due to the identity exchanges between the simulation boxes, which effectively decrease the difference between the estimated densities of the coexisting phases.
\par The order parameter behavior closely follows the asymptotic critical behavior with the Ising critical exponent $\beta$ = 0.326 in a surprisingly wide temperature range (solid lines in Fig.2). A more detailed analysis could be achieved by considering the temperature dependence of the effective exponent $\beta^{eff}$, that describes the behavior of the order parameter $\Delta\rho(\tau)$ at a given temperature (or temperature interval) and may be defined as
\begin{eqnarray}
\beta^{eff}(\tau) = d{\it ln}(\Delta\rho(\tau))/d{\it ln}\tau .
\end{eqnarray}
$\beta^{eff}$ can be derived by numerical differentiation of the  $\Delta\rho(\tau)$ or by analytic differentiation of Eq.(4) with the appropriate values of the fit parameters. We applied the latter method to plot the temperature dependence of $\beta^{eff}$ for the LJ fluid, studied in the present paper, and for some other model systems (see Fig.3). The deviations of the truncated LJ fluid from the asymptotic critical behavior (horizontal line) are the smallest among the available simulated systems. The full LJ fluid with long-range intermolecular potential shows strong positive deviations towards the mean-field critical behavior, which is characterized by the critical exponent $\beta$ = 0.50. The Ising model and square-well fluid with essentially short-range interactions show a strong crossover to a regular behavior, which corresponds to $\beta^{eff}$  $\rightarrow$ 0. So, the coexistence curve of the truncated LJ fluid shows really a very weak crossover both to the mean-field criticality and to the regular behavior and, therefore, it may be a suitable and promising model to study the surface critical behavior in a wide temperature range.
\begin{figure}[htb]
\includegraphics[width=8cm]{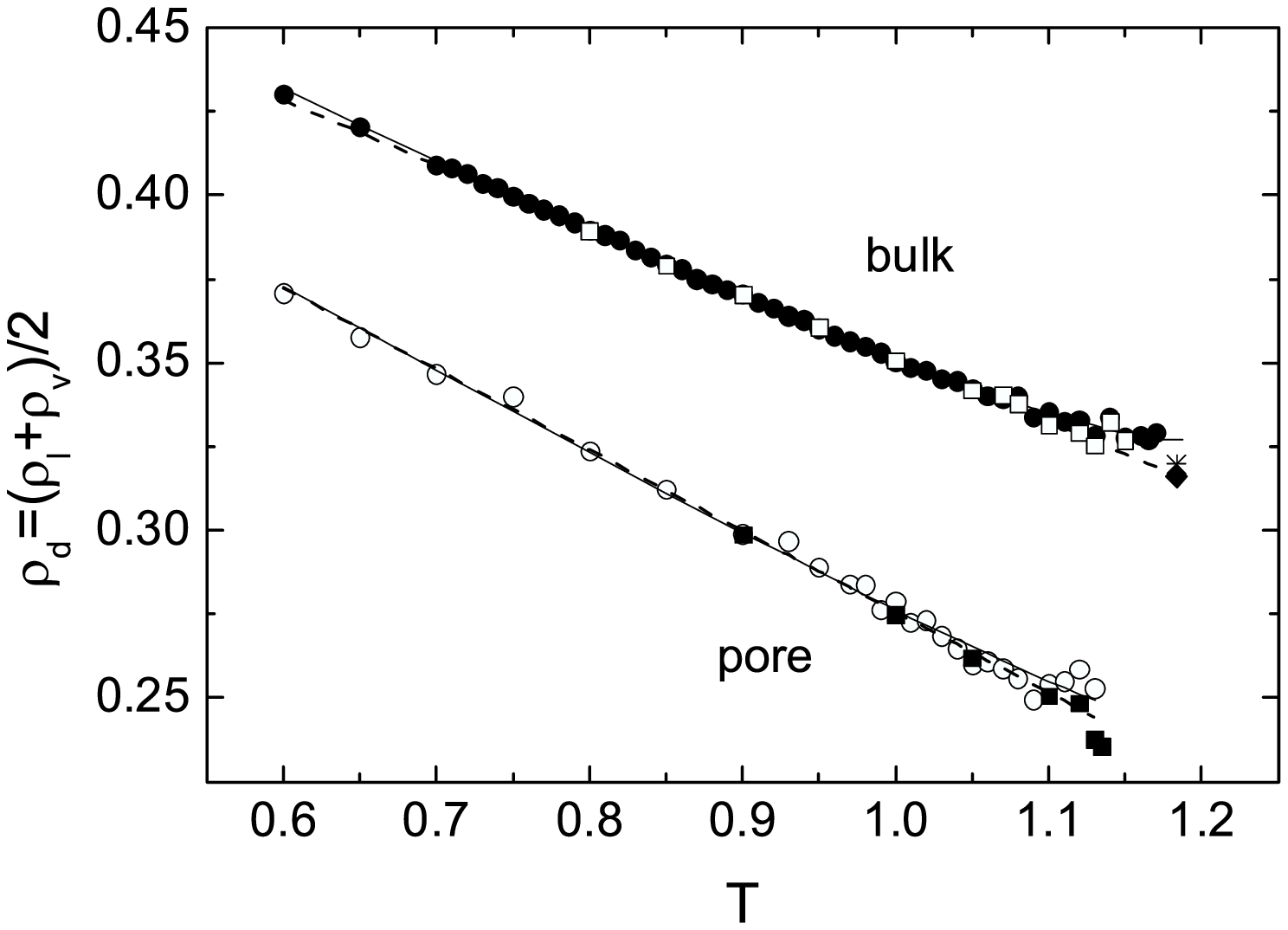}
\caption{Coexistence curve diameters for bulk LJ fluid with $\textit{L}$ $\approx$ 12 $\sigma$ (solid circles), bulk LJ fluid with L $\approx$ 20 $\sigma$ (open squares), LJ fluid in pore with H = 12 $\sigma$ and $\textit{L}$ $\approx$ 17 $\sigma$ (open circles) and LJ fluid in pore with H = 12 $\sigma$ and $\textit{L}$ $\approx$ 34 $\sigma$ (solid squares). The bulk critical densities, reported in Ref.\cite{Potoff98} (diamond) and Ref.\cite{Wilding95} (star) are shown for comparison. The linear fits 1 and 5 and fits 2 and 6 to Eq.(6) (Table III) are shown by dashed lines and solid lines, respectively.}
\end{figure} 
\par The diameter of the coexistence curve contains both regular contributions (critical density $\rho_c$, linear term, quadratic term, etc.) and singular terms ($\sim$ $\tau^{1-\alpha}$, $\sim$ $\tau^{2\beta}$, etc). Taking into account, that the singular term $\sim$ $\tau^{2\beta}$ could be very small in simple fluids \cite{FisherLut}, not extremely close to the critical point, the equation for the diameter can be written as
\begin{eqnarray}
\rho_d = (\rho_{l} + \rho_{v})/2 = \rho_{c}+B_1 \tau + B_{1-\alpha} \tau^{1-\alpha}+ ....
\end{eqnarray}
\begin{table}
\caption{Values of the parameters obtained from fits of Eq.(6) for the diameter $\rho_d$ to the data for the bulk and confined LJ fluid. $\chi^2$ represents the mean-square deviations with a confidence limit of 0.95 normalized to a dispersion of the simulated data of 0.001.}
\label{tab:3}       
\begin{tabular}{cccccc}
\hline\noalign{\smallskip}
N  & $\rho_c$ & B$_1$  & B$_{1-\alpha}$   & $\chi^{2}$ & comments\\ 
\noalign{\smallskip}\hline\noalign{\smallskip}
\multicolumn{6}{c}{bulk LJ fluid, L $\approx$ 12 $\sigma$} \\
1  & 0.3156(4) & 0.2277(15) & - & 1.23 & highest T = 1.11 \\
2   & 0.3269(7) & 0.647(32) & -0.403(30) & 1.15 & all data \\
\noalign{\smallskip}\hline\noalign{\smallskip}
\multicolumn{6}{c}{bulk LJ fluid, L $\approx$ 20 $\sigma$} \\
3   & 0.3182(11) & 0.2140(67) & -  & 5.2& all data \\
4   & 0.3131(35) & 0.544(221) & -0.30(20)  & 4.7& all data \\
\noalign{\smallskip}\hline\noalign{\smallskip}
\multicolumn{6}{c}{confined LJ fluid, H = 12 $\sigma$, $\textit{L}$ $\approx$ 17 $\sigma$} \\
5 & 0.2407(9) & 0.2764(41) & - & 5.6 & highest T=1.09 \\
6  & 0.2473(26) & 0.492(13) & -0.21(12) & 10 & all data\\
\noalign{\smallskip}\hline
\end{tabular}
\end{table}
The diameter of the simulated coexistence curve of the bulk LJ fluid is shown in Fig.4 and the various fits of the diameter values to Eq.(6) are collected in Table III. The diameter is essentially linear in a wide temperature range (see dashed line in Fig.4 and linear fits 1 and 3 in Table III) and only a few (4 or 7) points at the highest studied temperatures deviate upwards from the linear dependence. These points show clearly scattering around the average values, which give $\rho_c$ $\approx$ 0.335. The fits of the Eq.(6) to the diameter values reproduces a non-monotonic behavior of the diameter near T$_c$ (fits 2 and 4 in Table III, the former fit is also shown by a solid line in Fig.4). The value $\rho_c$ = 0.3269(7), obtained for the system with L $\approx$ 12 $\sigma$, noticeably exceeds the critical densities, reported in
 \cite{Wilding95,Potoff98,Trokhym99,Shi}. The value of the critical density, obtained from the GEMC simulations, could be underestimated due to the different numbers of particles in the two coexisting phases \cite{Vall,Vall1}. With increasing system size this effect becomes negligible and could shift the estimated critical density to higher values. However, our simulations show the opposite trend: increasing the system size to $\textit{L}$ $\approx$ 20 $\sigma$ causes a decrease of the critical density to $\rho_c$ = 0.3182(11) in the case of a linear fit (fit 3 in Table III) or even to $\rho_c$ = 0.3131(35) by fitting Eq.(6) (fit 4 in Table III). So, an increase of the system size from $\textit{L}$ $\approx$ 12 $\sigma$ to $\textit{L}$ $\approx$ 20 $\sigma$ shifts the value of the critical density closer to the values $\rho_c$ = 0.316(3)\cite{Potoff98} and $\rho_c$ = 0.3197(4)\cite{Wilding95}, obtained by finite-size scaling methods. Note also, that the mixed-field scaling method, used in \cite{Wilding95,Potoff98}, overestimates the value of $\rho_c$ of the square-well fluid due to the neglect of the Yang-Yang anomaly \cite{Fishermethod}.
\begin{figure}
\includegraphics[width=8cm]{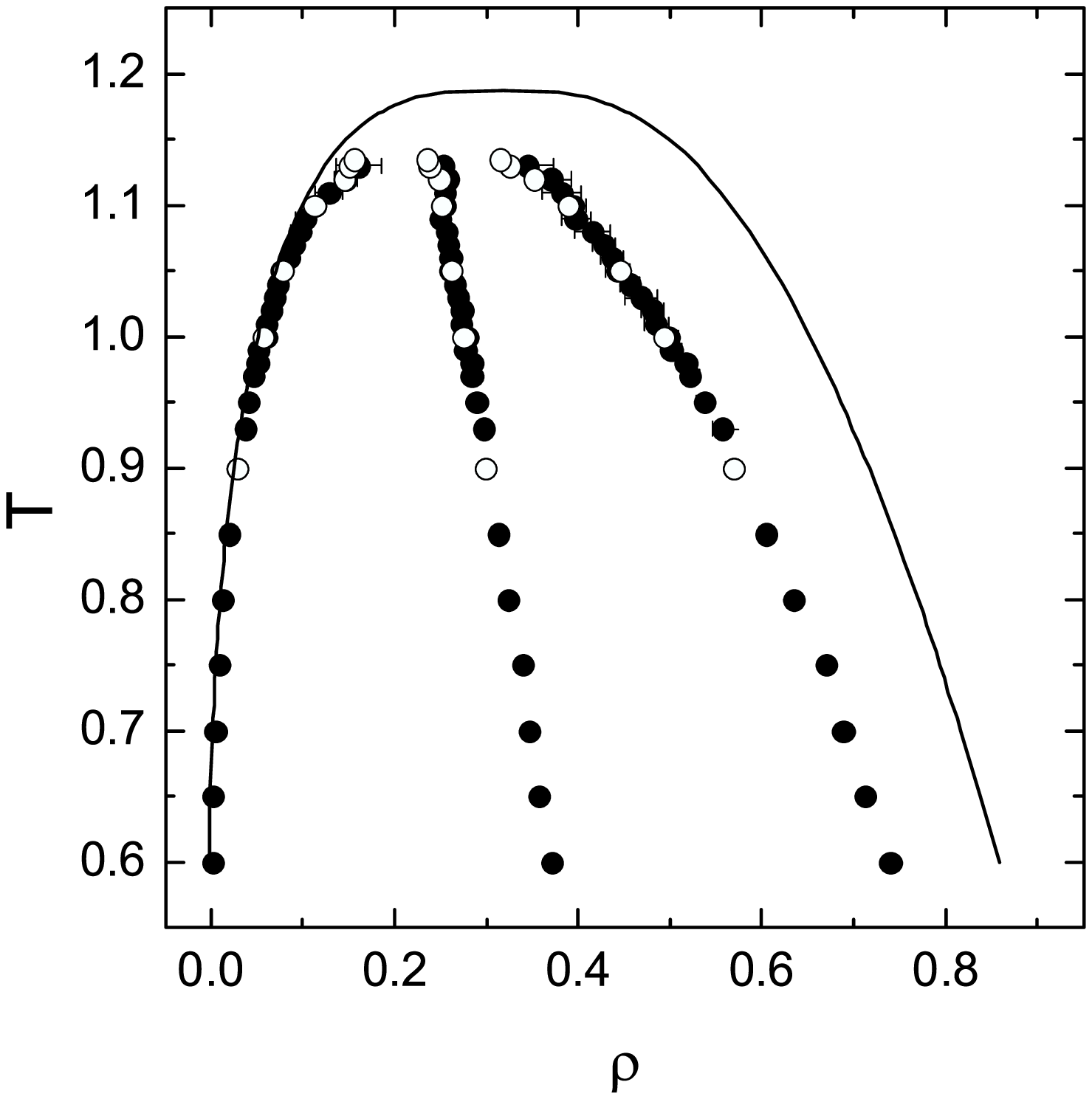}
\caption{Liquid-vapor coexistence curve and diameter of the LJ fluid in the slitlike pore H = 12 $\sigma$, L $\approx$ 17 $\sigma$ (solid circles) and H = 12 $\sigma$, L $\approx$ 34 $\sigma$ (open circles). The bulk coexistence curve (line) is reproduced using Eq.(4) with set 4 of the fitting parameters from Table II and a linear diameter (fit 1 in Table III).}
\end{figure}
\subsubsection{Pore coexistence curve}
\par
The average density of the LJ fluid confined in a pore was calculated assuming that the pore volume is equal to $\textit{L$^2$H}$. In fact, the fluid-wall interaction is equal to zero at the distance 0.86 $\sigma$ from the pore wall and so in an operational approach this interval could be divided equally between the volumes of the fluid and the solid. As a result, the real volume accessible to the fluid molecules should be reduced by a factor 1.077 and the average pore density should increase accordingly. Throughout the paper we use the average density in the pore without this correction on the accessible volume.
\par
The simulated coexistence densities and the diameter for the LJ fluid in the slitlike pore with weakly attractive walls of width H = 12 $\sigma$ are shown in Fig.5. The data points obtained from the simulations using the pore with lateral size $\textit{L}$ $\approx$ 17 $\sigma$ and the pore with lateral size $\textit{L}$ $\approx$ 34 $\sigma$ coincide,  except at the 3 highest temperatures, where the density of the liquid phase becomes lower in the large system (see Fig.5). The critical temperature of the pore coexistence curve is depressed due to the effect of the confinement. The vapor phase in the pore has a larger density than in the bulk and this effect increases with temperature. At low temperatures this may be understood as a result of adsorption at the weakly attractive substrate. At high temperatures this effect is the result of a higher coexistence pressure due to the  confinement in a pore with weakly attractive substrate \cite{Evans}. The density of the liquid phase in the pore is shifted significantly to lower values due to a depletion of the density near the weakly attractive surface \cite{BGO2004} and this effects the liquid density much stronger than the increasing pressure.
\par The pore critical temperature is located between the highest temperature, where a two-phase coexistence was obtained (T = 1.13) and the lowest temperature, where the two phases become identical in the GEMC simulations (T = 1.15). In a slitlike confinement the 3D critical behavior is distorted by a temperature crossover towards two-dimensionality when approaching the pore critical temperature and by the spatial heterogeneity of the fluid density at all temperatures due to the influence of the surface. There are no theoretical equations which give a description of the pore coexistence curve. Fits of the Eq.(4) to the pore coexistence curve show, that the critical temperature varies slightly depending on whether the critical exponent $\beta$ is fixed to the 2D or 3D value (see Table II). We determine for the pore critical temperature the value T$_c$ = 1.145(2). 
\par
The diameter of the pore coexistence curve is shown in Fig.4. A linear temperature dependence is observed in a wide range as in the bulk case (see dashed lines). The fits of the diameter to Eq.(6) are given in Table III (fits 5 and 6). Deviations of the high-temperature points upwards from the linear dependence and convergence to the value $\rho_c$ $\approx$ 0.255 can be noticed. In the larger simulated system with L $\approx$ 34 $\sigma$ this effect disappears (see Fig.4).
\begin{figure}
\includegraphics[width=8cm]{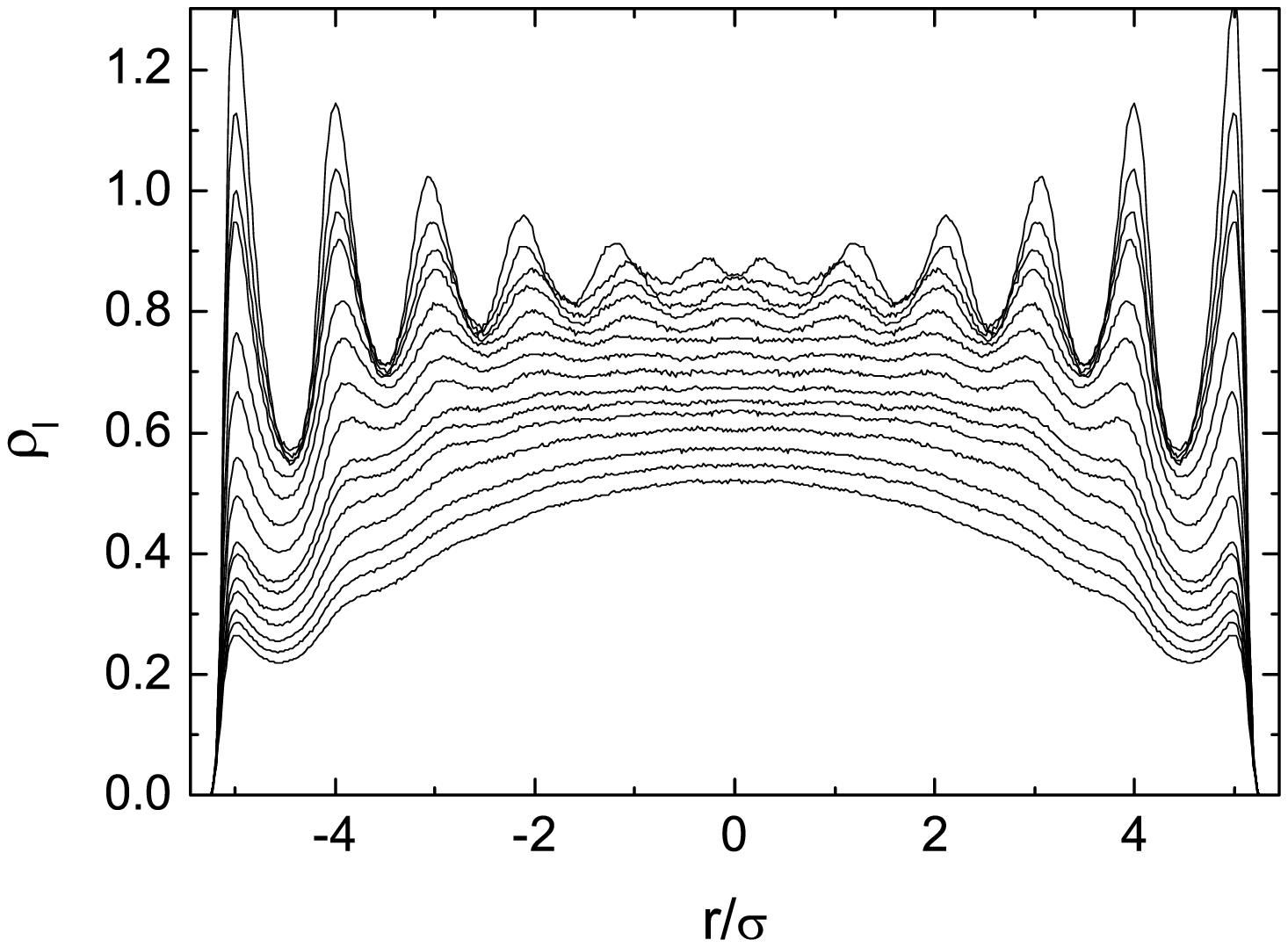}
\caption{Density profiles of the LJ liquid in a slitlike pore with H = 12 $\sigma$ in equilibrium with a coexisting vapor phase. The temperature increases from the top (T = 0.60) to the bottom (T = 1.13).}
\end{figure}
\begin{figure}
\includegraphics[width=8cm]{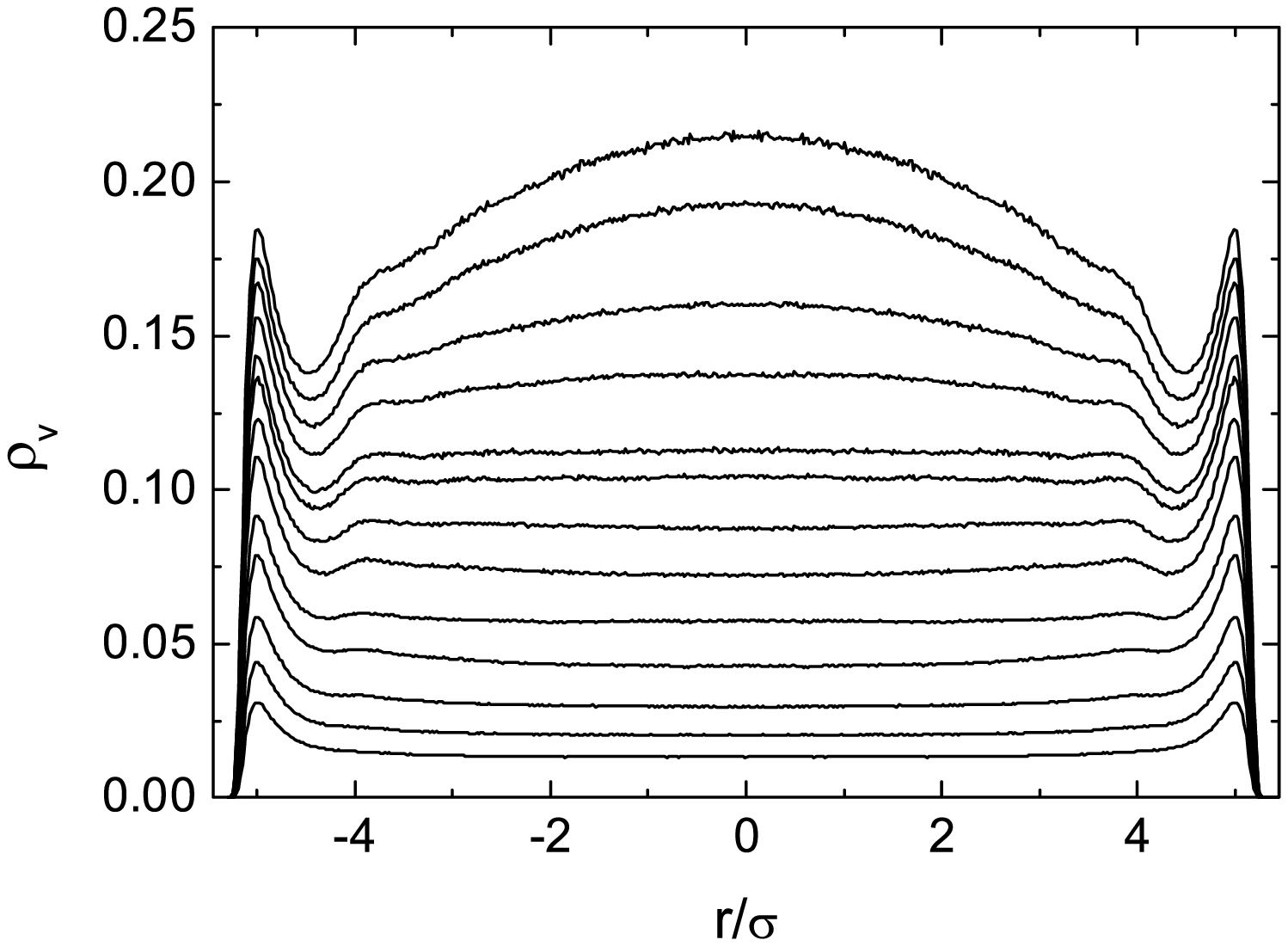}
\caption{Density profiles of the LJ vapor in the slitlike pore H = 12 $\sigma$ in equilibrium with the coexisting liquid phase. Temperature increases from the bottom (T = 0.80) to the top (T = 1.13).}
\end{figure}
\subsubsection{Density profiles}
The density of the coexisting liquid phase in the pore is significantly lower than the coexisting liquid density in the bulk (Fig.5). This originates from a depletion of the liquid density near the weakly attractive substrate and can be studied by an analysis of the density profiles $\rho_{l,v}$(r), where r is the distance from the pore center. At low temperatures the density profiles in the liquid phase show strong oscillations due to packing effects, which are still pronounced even near the pore center (Fig.6, upper curves). Such strong density oscillations prevent observing a density depletion at the temperatures T = 0.60 to T = 0.80. With increasing temperature the density profiles become more smooth. However, first and second shell oscillations are still to be seen even at the highest temperature T = 1.13 (Fig.6, lowest curve). Visual examination of the liquid profiles in Fig.6 evidences that the density near the surface decreases faster with temperature than the density in the pore interior.  
\begin{figure}
\includegraphics[width=8cm]{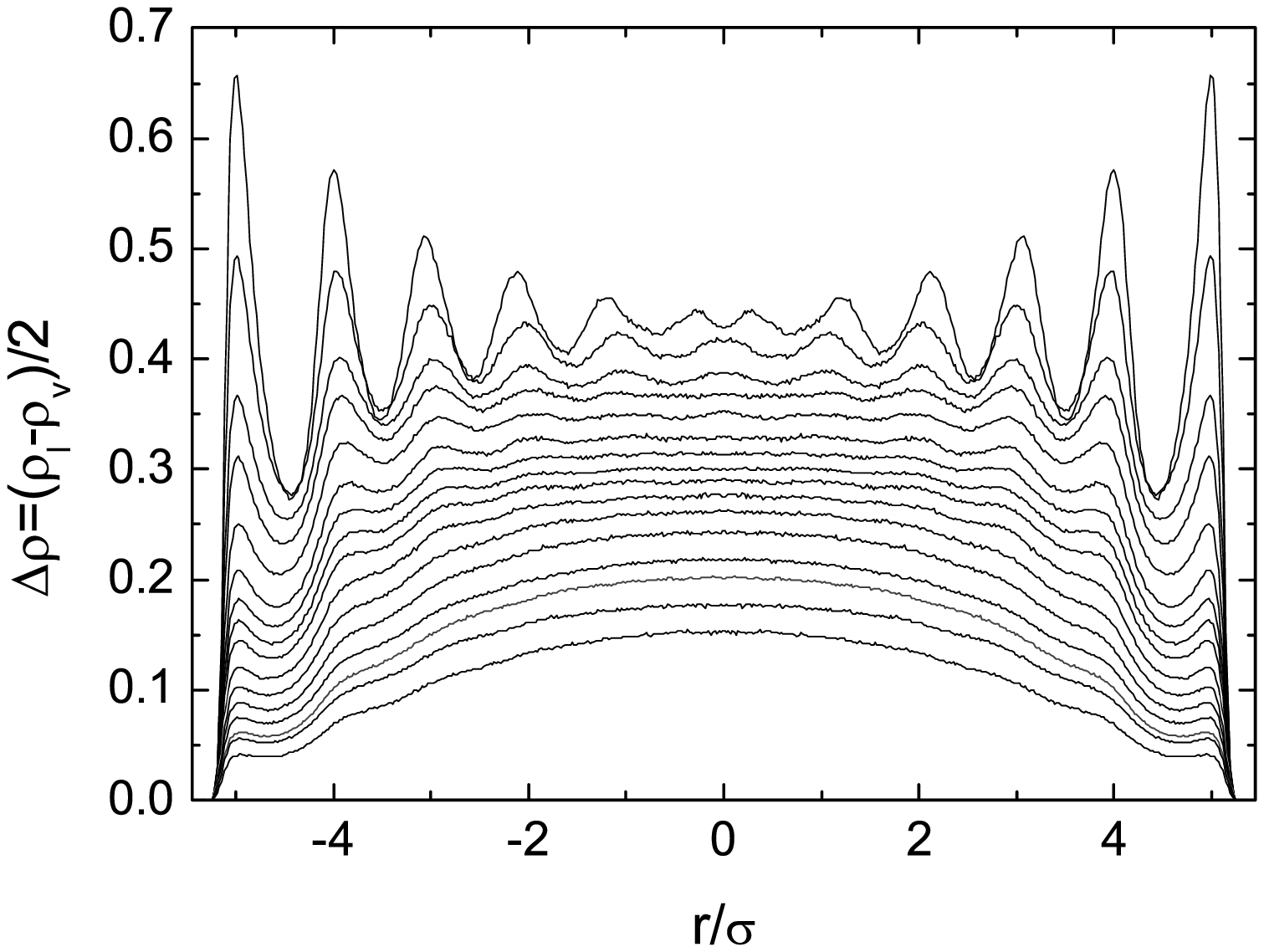}
\caption{Profiles of the order parameter for the LJ fluid in the slitlike pore with H = 12 $\sigma$. Temperature increases from the top (T = 0.60) to the bottom (T = 1.13).}
\end{figure} 
\par
The temperature evolution of the density profiles of the vapor phase is shown in Fig.7. The vapor density maximum near the wall remain pronounced in the whole temperature range studied. Its position coincides with the first density maximum in the liquid phase profiles and corresponds to the minimum of the fluid-wall potential at a distance $\sigma$ from the pore wall (see Eq.(3)). In the pore interior the vapor density varies smoothly and remains flat up to T = 1.07. At higher temperatures the profile of the vapor becomes convex upwards and a second density maximum can be seen at the highest temperatures due to the appearance of packing effects with increasing fluid density (Fig.7).  
\par 
To study the surface critical behavior, we define the local order parameter $\Delta\rho$($\textit{r}$,$\tau$) and the local diameter $\rho_d$($\textit{r}$,$\tau$) of the fluid near a surface similarly to the bulk case
\begin{eqnarray}
\Delta\rho(r,\tau) = (\rho_l(r,\tau)-\rho_v(r,\tau))/2\\
\rho_d (r,\tau) = (\rho_l(r,\tau)+\rho_v(r,\tau))/2
\end{eqnarray}
The profiles of the order parameter $\Delta\rho$($\textit{r}$,$\tau$) are shown in Fig.8. Due to a partial compensation of the density oscillations in the liquid and vapor phases, the order parameter profiles are smoother than the liquid or vapor profiles. 
\par
\begin{figure}
\includegraphics[width=8cm]{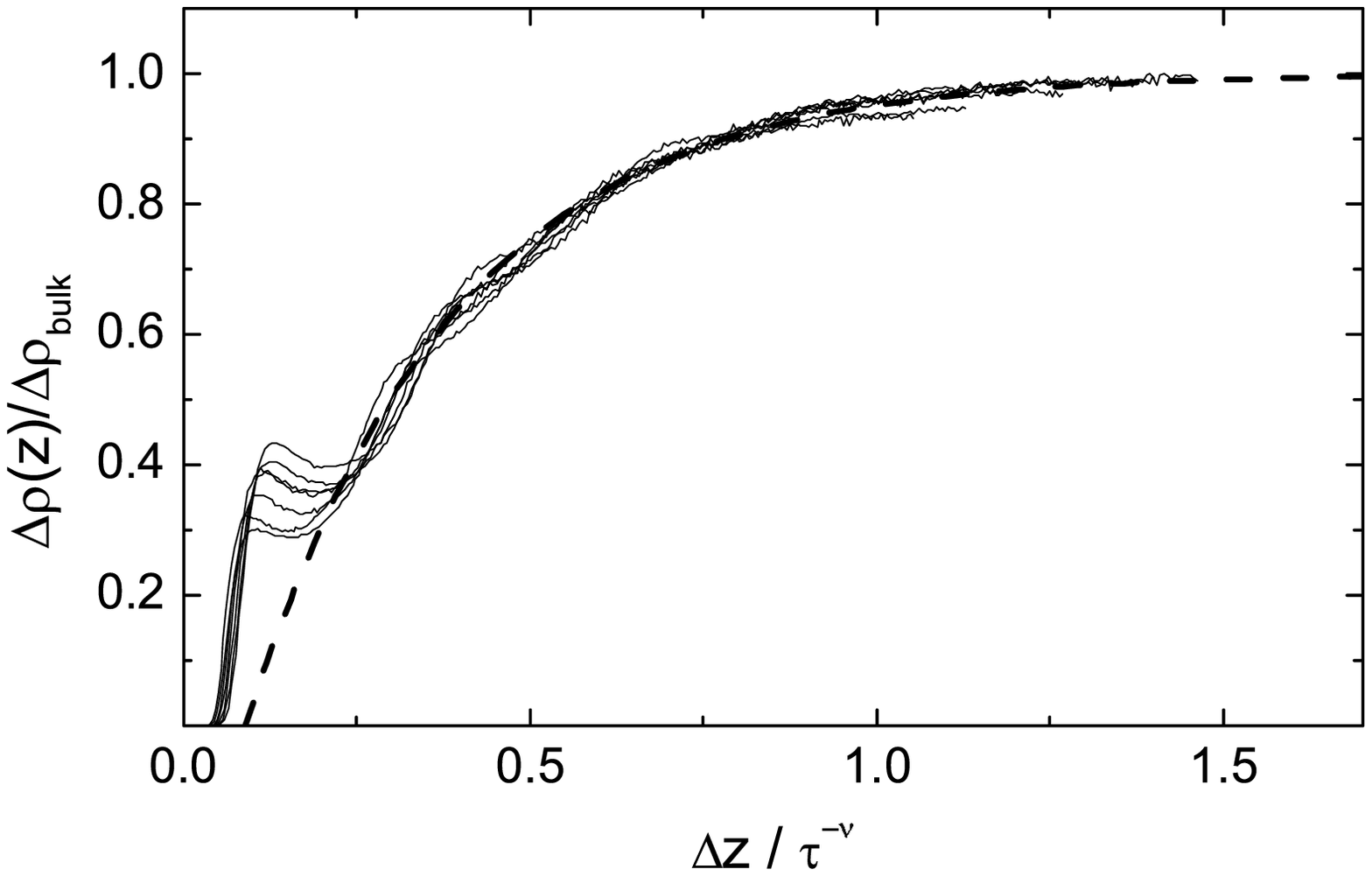}
\caption{Scaling of the order parameter profiles for seven temperatures from T = 1.04 to 1.10. The dashed line is the function $\Delta\rho$($\Delta$z)/$\Delta\rho_{bulk}$ = 1 - 
exp(-($\Delta$z - 0.09 $\sigma$)/$\xi_0 \tau^{-\nu}$) with $\nu$ = 0.63, $\xi_0$ = 0.30 $\sigma$. For definition of $\Delta$z see text.}
\end{figure}
The profile of the order parameter in a lattice system is free from oscillations near the boundary and may be described by smooth functions. At distances larger than the bulk correlation length $\xi$, the deviation of the order parameter from the bulk value decays exponentially, when moving away from the surface \cite{Binderrev1}. Near the surface (at distances $\Delta z$ $<<$ $\xi$) the order parameter profile should obey an algebraic law. Due to the density oscillations of the fluid near the structureless surface the algebraic law could be observed only in close proximity to the critical point, when the bulk correlation length significantly exceeds the range of density oscillations. In a confined fluid the approach to the bulk critical temperature is limited by the pore critical temperature and therefore the algebraic region may not be detected. That is why we fitted the order parameter profiles to a simple exponential law \cite{Binderrev1}:
\begin{eqnarray}
\Delta\rho(\Delta z,\tau) = \Delta\rho_{bulk}(\tau)\left[1 - exp\left(-\frac{\Delta z - \lambda}{\xi_-}\right)\right],
\end{eqnarray}
where $\Delta\rho_{bulk}$($\tau$) is the order parameter of the bulk fluid; $\xi_-$ is the bulk liquid orrelation length along the coexistence curve. Taking into account, that distances closer then about 0.5 $\sigma$ to the wall are not accessible to the centers of particles, we define $\Delta$z as the distance to a parallel plane 0.5 $\sigma$ inside the pore. As this definition contains some ambiguity, we introduced in Eq.(9) the adjustable parameter $\lambda$. 
\par
First, we examine whether $\Delta\rho$($\Delta$z,$\tau$) really follows the universal critical behavior, described by Eq.(9). Therefore we plot $\Delta\rho$($\Delta$z,$\tau$)/$\Delta\rho_{bulk}$($\tau$) vs the normalized distance $\Delta$z/$\xi_-$ with $\xi_-$ = $\xi_0\tau^{-\nu}$, $\nu$ = 0.63. In the temperature range 1.04 $<$ T $<$ 1.10 a universal master curve perfectly describes the order parameter profiles, excluding the first oscillation near the wall (see Fig.9). The dashed line in Fig.9 corresponds to an amplitude of the correlation length $\xi_0$ = 0.30 $\sigma$ and $\lambda$ = 0.09 $\sigma$. Note, that the center of the first fluid layer is situated near the potential minimum at $\Delta$z = 0.5 $\sigma$, while the fluid-wall interaction is repulsive at $\Delta$z $\leq$ 0. So, the order parameter becomes equal to zero at some distance between the minimum and the zero value of the fluid-wall potential. At higher temperatures (T $>$ 1.10, not shown in Fig.9) the order parameter profiles $\Delta\rho$($\Delta$z,$\tau$) deviate in the pore interior from the master curve to lower values. This effect is caused by the influence of the opposite wall and will be discussed below. At lower temperatures (T $<$ 1.04, not shown in Fig.9) the oscillations of the order parameter become more pronounced. This limits the temperature range where the validity of the universal behavior could be analyzed. 
\begin{figure}
\includegraphics[width=7cm]{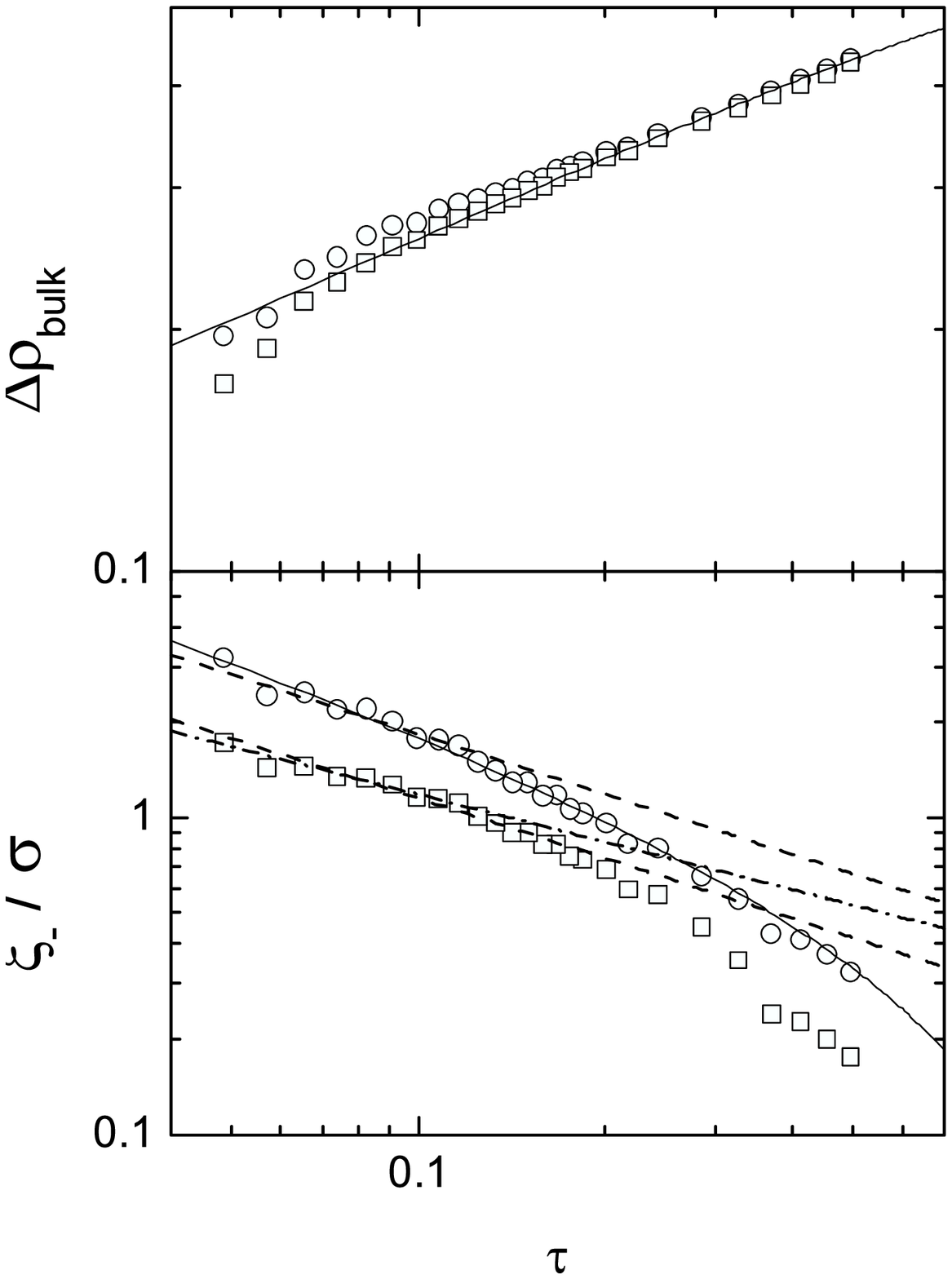}
\caption{Temperature dependence of the fitting parameters $\Delta\rho_{bulk}$ (upper panel) and $\xi_-$ (lower panel), obtained from fits of Eq.(9) (open circles) or Eq.(10) (open squares) to $\Delta\rho$($\Delta$z,$\tau$). The solid line in the upper panel shows the behavior of the bulk order parameter. In the lower panel the dashed line has the slope $\nu$ = 0.63, the dot-dashed line has the mean-field slope $\nu$ = 0.5. The solid line in the lower panel is the fitting curve $\xi_-$ = 0.57$\tau^{-\nu}$ (1 - 0.89$\tau^{0.52}$)$\sigma$.}
\end{figure}
\begin{figure}
\includegraphics[width=8cm]{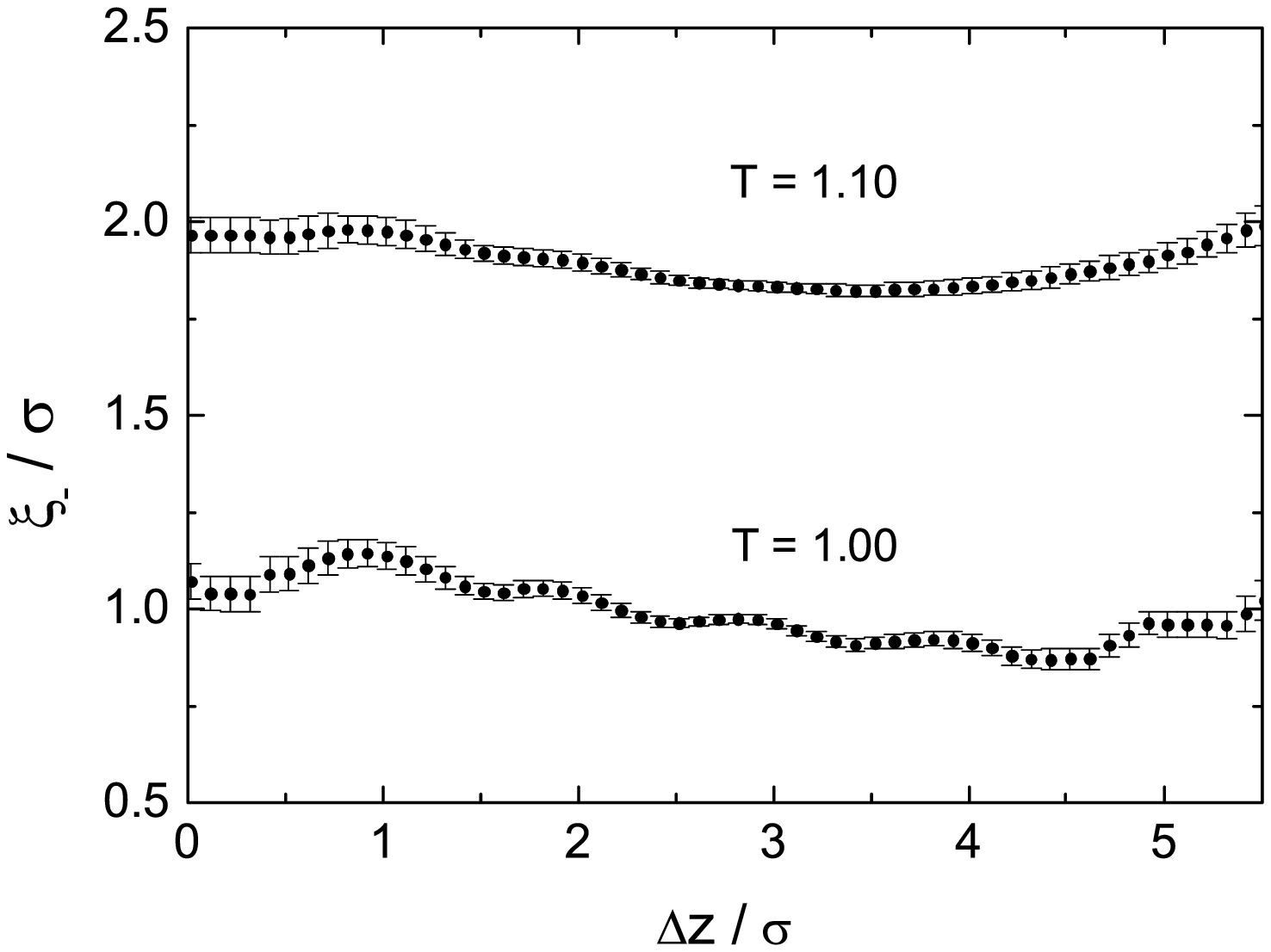}
\caption{Variation of the fitting parameter $\xi_-$, when Eq.(9) is fitted to parts of the order parameter profiles $\Delta\rho$($\Delta$z,$\tau$), which extend from the wall to various distance $\Delta$z from the surface (with fixed $\lambda$ = 0). }
\end{figure}
\par 
In a second approach, the correlation length was estimated from fits of Eq.(9) to each profile $\Delta\rho$($\Delta$z,$\tau$). The obtained values of $\Delta\rho_{bulk}$ and $\xi_-$ are shown in Fig.10. The parameter $\lambda$ fluctuated around zero and never exceeded a few hundredths of $\sigma$, when the whole profile $\Delta\rho$($\Delta$z,$\tau$) is used for fitting. If only parts of the profile $\Delta\rho$($\Delta$z,$\tau$) are used and the parameter $\lambda$ is fixed to 0, the obtained values of $\Delta\rho_{bulk}$ and $\xi_-$ varied no more than about 5 to 10 $\%$. For illustration, the values $\xi_-$, obtained from fits to parts of the order parameter profiles extending from the wall to various distances $\Delta$z are shown in Fig.11 for two temperatures.
 Comparison of the values $\Delta\rho_{bulk}$, which were obtained from the fitting of Eq.(9), with the values of the true bulk order parameter, obtained from the bulk coexistence curve (Fig.10, upper panel), shows, that the two sets coincide only at low temperatures. With increasing temperature the fitted values of $\Delta\rho_{bulk}$ deviate slightly upward from the true bulk data.  At the highest 3 to 5 temperatures the fitted values of $\Delta\rho_{bulk}$ rapidly turn down, which correlates with the deviations of the $\Delta\rho$($\Delta$z,$\tau$) profiles at these temperatures from the master plot, shown in Fig.9. At the highest temperatures the temperature dependence of the correlation length $\xi_-$ agrees with the expected asymptotic critical behavior with an amplitude $\xi_0$ = 0.43 $\sigma$ (Fig.10, lower panel, dashed line), while at lower temperatures it bends downward. Note, that the whole data set of $\xi_-$ can be fitted by an equation with one correction-to-scaling term (see solid line at the lower panel of Fig.10), using an amplitude $\xi_0$ = 0.57 $\sigma$ and a comparatively large amplitude (0.89) of the correction-to-scaling term. Thus the values $\xi_0$ = 0.43 $\sigma$ and $\xi_0$ = 0.57 $\sigma$ of the amplitude of the correlation length were obtained when fitting the order parameter profiles including the first oscillation. Naturally, this effectively causes an increase of the obtained values $\xi_0$. An estimate of $\xi_0$ from the universal master curve (Fig.9) neglecting the first oscillation of the order parameter results in the lower value $\xi_0$ = 0.30 $\sigma$. As the first density oscillation reflects only some non-universal microscopic detail of the interaction potential, the latter estimate seems to be more reasonable. For comparison, the only estimate of the correlation length for a LJ fluid (truncated, however, at another radius) from molecular dynamic simulations gives an amplitude of the correlation length $\xi_0$ $\sim$ 0.27 $\pm$ 0.2 $\sigma$\cite{xiLJ}, using the critical temperature of the LJ fluid with full potential \cite{Lotfi}. However, use of the critical temperature of the LJ fluid with truncated potential \cite{xiLJ} increases the  value $\xi_0$ almost by a factor of two. The available experimental studies of noble gases give for subcritical temperatures values of $\xi_0$ of about 20 to 22$\%$ of their Van-der-Waals diameters \cite{Sengers}. 
\par
The values $\xi_-$, which were obtained from the local order parameter profiles of a fluid in a slitlike geometry, differ from the bulk correlation lengths in semi-infinite systems due to two main effects. The shift of the liquid-vapor phase transition due to the confinement decreases $\xi_-$, whereas the effect of the opposite wall could increase the "effective" value of the estimated correlation length. Thus, a non-monotonous variation of the "effective" value of $\xi_-$ with the pore width could be expected. To study these effects, simulations of the coexisting densities in pores of larger width should be performed and these studies are in progress now. This should provide a new way to study bulk correlation lengths of fluids by computer simulations.  
\par Additionally we have fitted the order parameter profile by the mean-field equation \cite{Binderrev1}:
 \begin{eqnarray}
\Delta\rho(\Delta z,\tau) = \Delta\rho_{bulk}(\tau)tanh\left(\frac{\Delta z-\lambda}{2\xi_-}\right).
\end{eqnarray}
These fits give values of the correlation length $\xi_-$ which are about 1.6 times lower than the values, obtained from the fits using Eq.(9) (see Fig.10). At high temperatures the behavior of the correlation length $\xi_-$ is compatible both with the Ising model ($\nu$ = 0.63, dashed line in fig.10) and with the mean-field prediction ($\nu$ = 0.5, dot-dashed line in Fig.10). The behavior of $\Delta\rho_{bulk}$ obtained from the fits using the mean-field Eq.(10) agrees well with the data for the bulk coexistence curve, excluding the tree highest temperatures (see Fig.10, upper panel). 
\subsubsection{Thermal and spatial crossover from bulk-like to surface-like critical behavior}
\begin{figure}
\includegraphics[width=8cm]{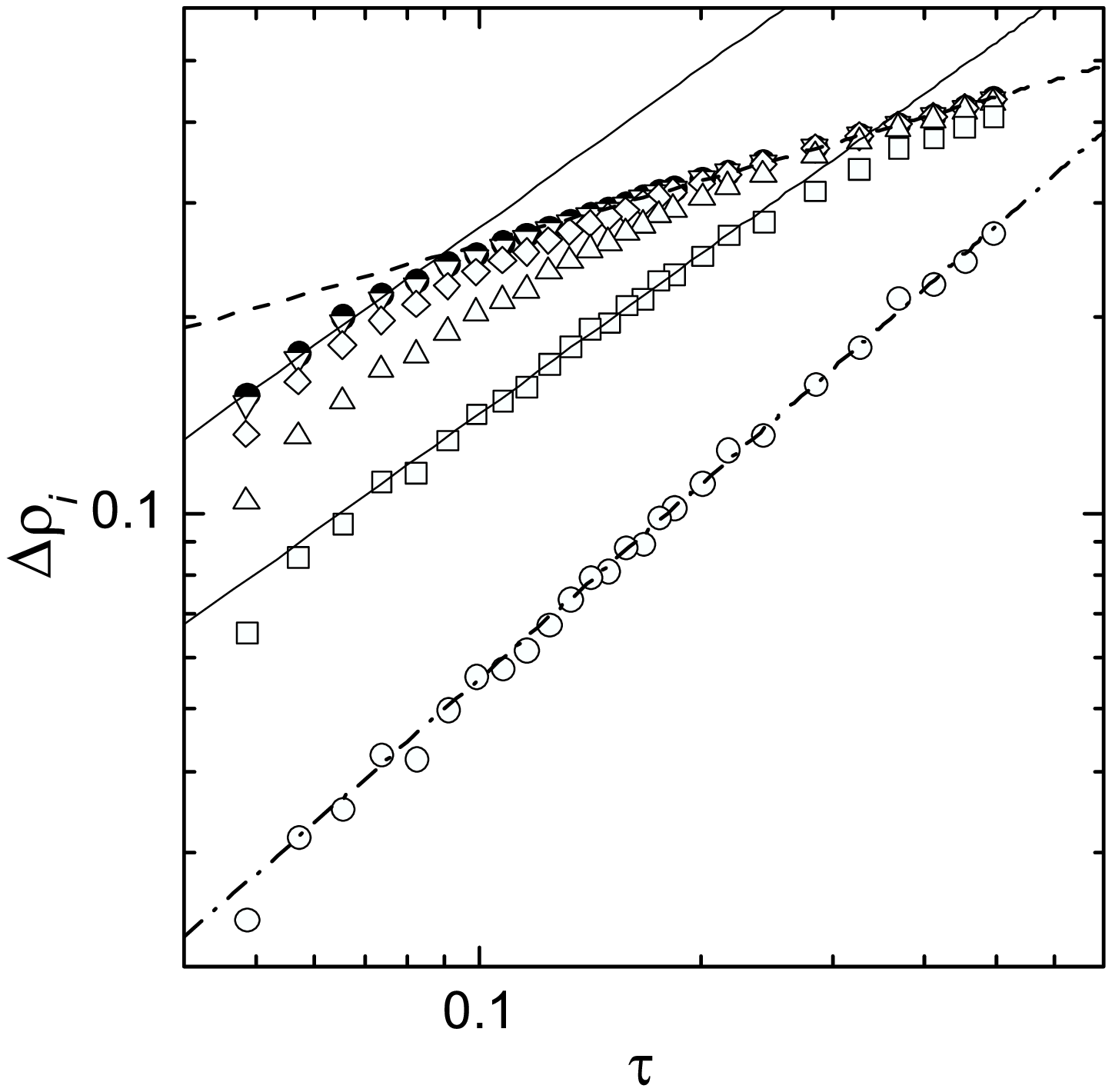}
\caption{A double logarithmic plot of the local order parameter, averaged over layers of molecular width (see text for the details) vs reduced temperature: first (surface) layer (open circles), 2-nd to 5-th layers (squares, triangles, diamond, triangle down, respectively); 6-th layer (pore interior, solid circles). The slopes of the lines correspond to the exponent $\beta$ = 0.326 (dashed line), $\beta_1$ = 1 (dot-dashed line), $\beta_1$ = 0.82 (solid lines).}
\end{figure}
Density profiles of the liquid and vapor phases at the pore coexistence curve were used to study the temperature dependence of the local order parameter $\Delta\rho$($\Delta$z,$\tau$). A double logarithmic plot of the order parameter $\Delta\rho_i$($\tau$), averaged over layers of about molecular width $\sigma$ vs the reduced temperature $\tau$ is shown in Fig.12. The layer between the zero energy point of the fluid-wall interaction (0.43 $\sigma$ from the wall, $\Delta$z = -0.07) and the first minimum in the density oscillations (1.5$\sigma$ from the wall, $\Delta$z = 1) was defines as the first (surface) layer. Each subsequent layer spreads over 1 $\sigma$ toward the pore center. The order parameter in the surface layer $\Delta\rho_1$($\tau$) changes practically linearly with $\tau$ (dot-dashed line in Fig.12). In the second layer the behavior of $\Delta\rho_2$($\tau$) is close to a power law with exponent $\beta_1$ = 0.82 in a wide range of temperatures (see squares and solid line in Fig.12). In the third and subsequent layers a crossover from bulk-like behavior (dashed line, Fig.12) to surface behavior at smaller $\tau$, i.e. closer to the critical point is observed.
\par
\begin{figure}
\includegraphics[width=8cm]{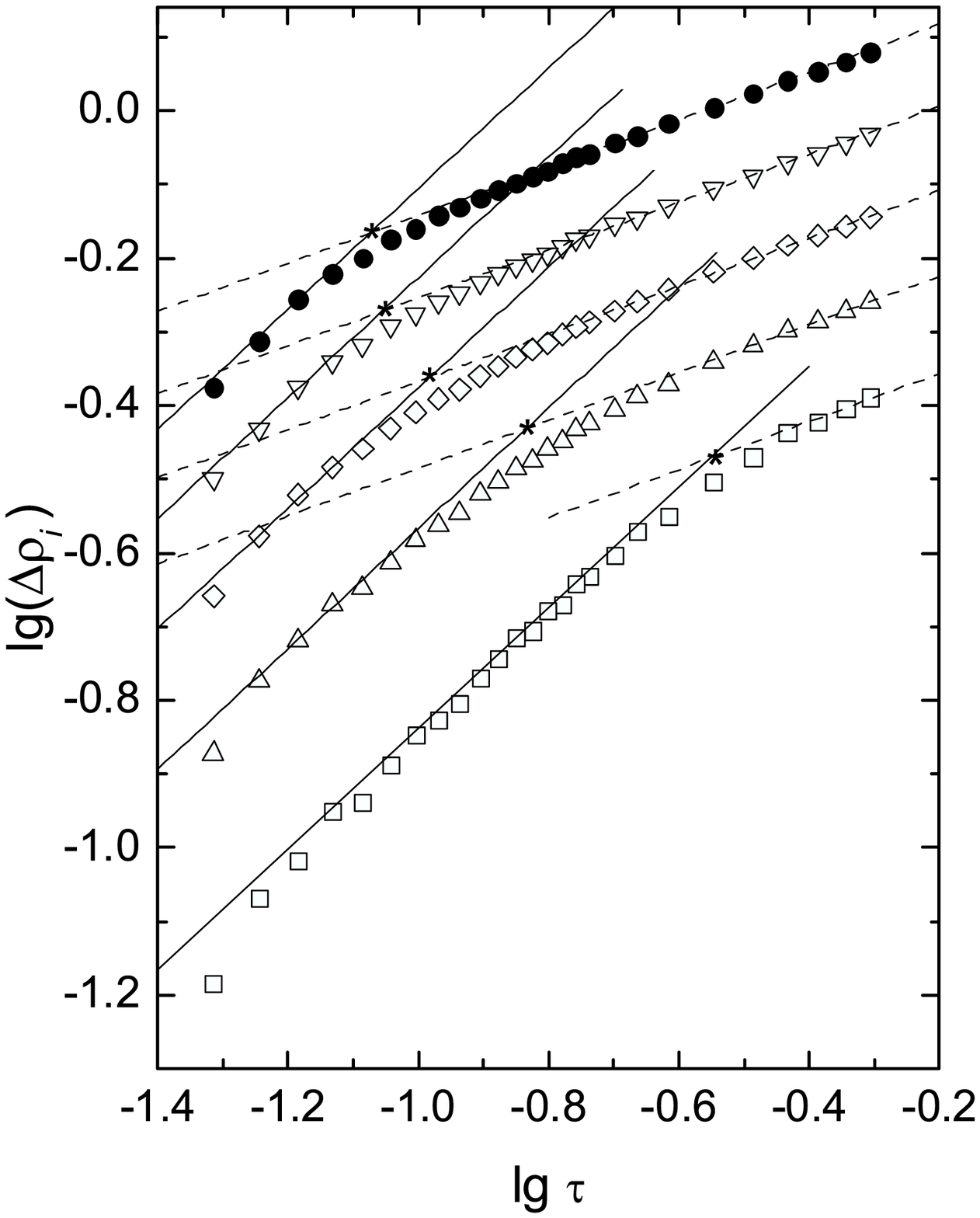}
\caption{Lines and symbols are the same as in Fig. 12. The data for the 3-rd and subsequent layers are shifted vertically. The position of the crossover points from bulk-like to surface-like behavior in each layer are denoted by stars. }
\end{figure}
The crossover temperature could be extracted from this plot as the temperature of the crossing point of the two straight lines, which represent bulk behavior with the exponent $\beta$ = 0.326 (dashed lines in Fig.13) and surface critical behavior with the exponent $\beta$ = 0.82 (solid lines). These crossover points are shown by stars in Fig.13 for the 2-nd to 6-th molecular layers. 
\begin{figure}
\includegraphics[width=8cm]{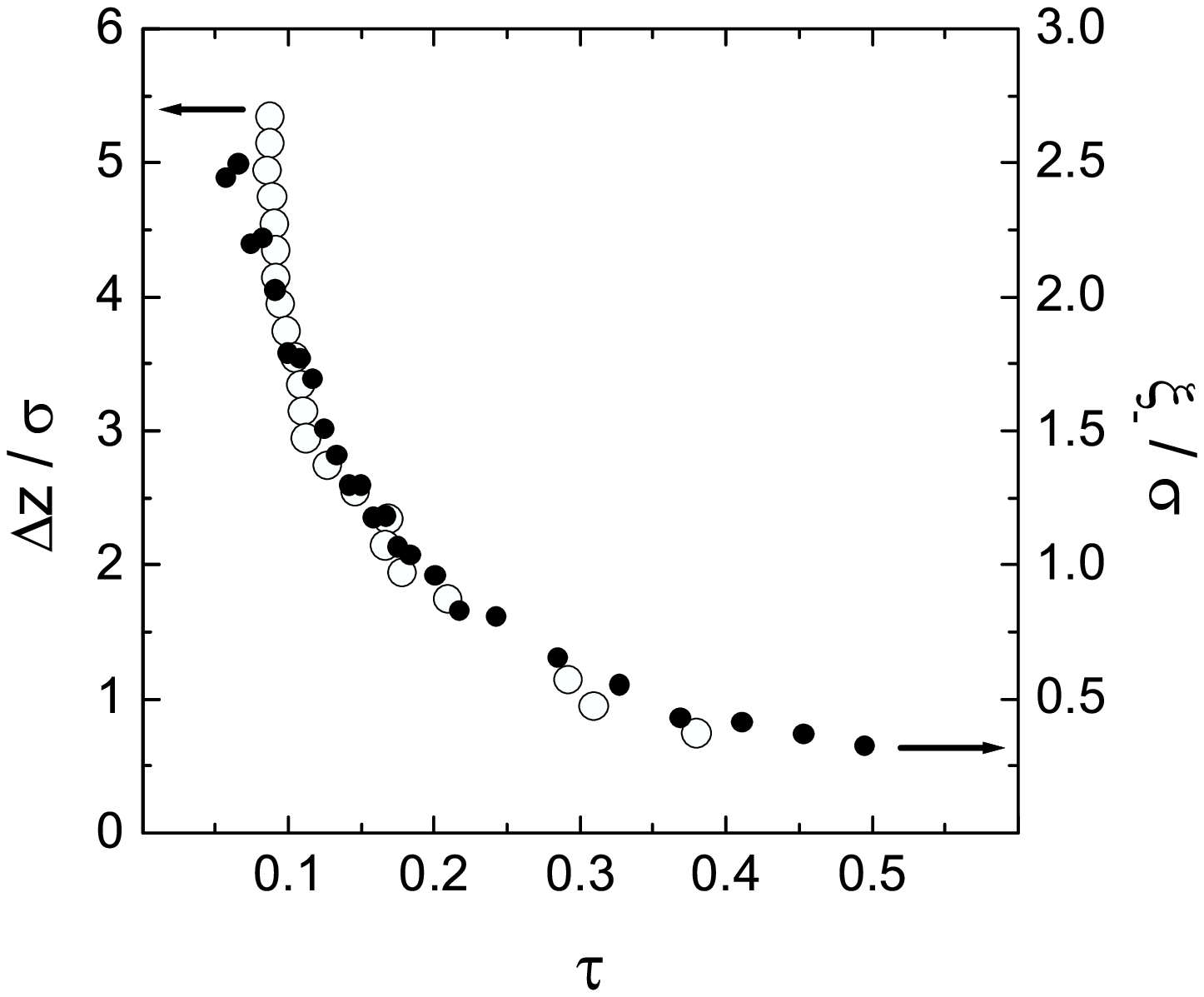}
\caption{Layer distance $\Delta$z vs crossover temperature (left axis, open symbols) and temperature dependent correlation length $\xi_-$($\tau$) found from fits of Eq.9 to the profiles $\Delta\rho$($\Delta$z,$\tau$) (right axis, solid circles).}
\end{figure}
\par
To study the dependence of the corresponding crossover temperature $\tau^*$ on the distance from the surface  $\Delta$z in more detail, we performed the same analysis for a finer grid of layers of 0.2 $\sigma$ width. The obtained dependence is shown in Fig.14. The values of the correlation length $\xi_-$, obtained from fits of Eq.(9) to the order parameter profiles $\Delta\rho$($\Delta$z,$\tau$), are also shown in Fig.14 (right axis, solid circles). Comparison of the two dependencies shown in Fig.14 evidences that the crossover from bulk to surface critical behavior is governed by the correlation length: the crossover in a definite fluid layer occurs, when the correlation length $\xi_-$ corresponds roughly to half the distance from the layer to the solid surface: $\Delta$z $\approx$ 2$\xi_-$. 
\begin{figure}
\includegraphics[width=8cm]{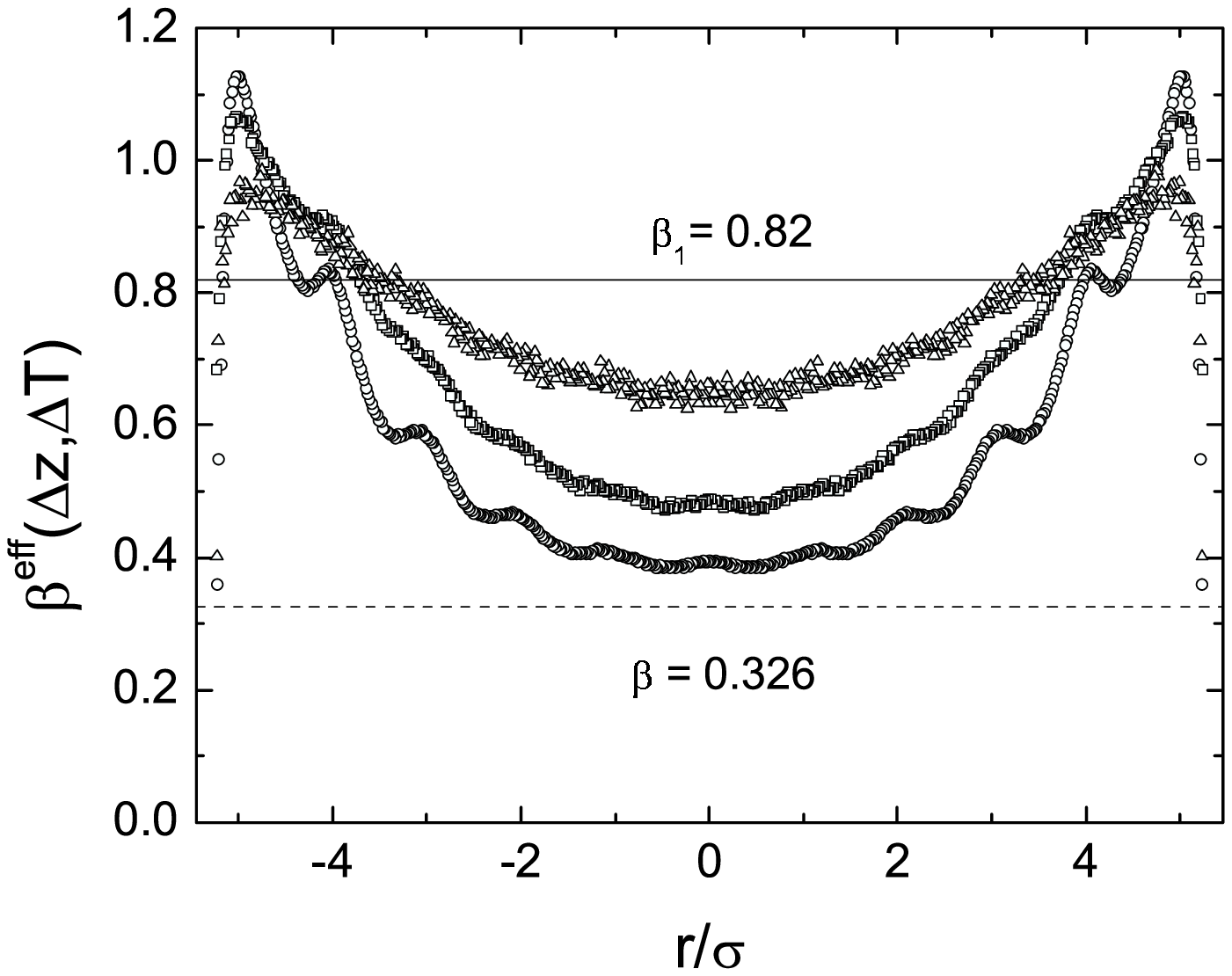}
\caption{Spatial crossover of the local effective exponent $\beta^{eff}$($\Delta$z, $\Delta$T), determined from Eq.(11) applied to the local order parameter $\Delta\rho$($\Delta$z,$\tau$) in different temperature intervals $\Delta$T: 0.72 $\leq$ T $\leq$ 1.12 (circles, lowest curve), 1.01 $\leq$ T $\leq$ 1.12 (squares, middle curve) and 1.08 $\leq$ T $\leq$ 1.12 (triangles, upper curve). The critical exponent of the bulk fluid $\beta$ and the surface critical exponent $\beta_1$ of the Ising model are shown by dashed and solid horizontal lines, respectively. r is the distance from the pore center, $\Delta$z the distance from the wall.}
\end{figure} 
\par 
The spatial crossover from bulk to surface critical behavior was also studied by using the effective local critical exponent $\beta^{eff}$($\Delta$z,$\Delta$T), which was obtained from fitting the temperature dependence of $\Delta\rho$($\Delta$z,$\tau$) by the equation
\begin{eqnarray}
\Delta\rho(\Delta z,\tau) = B^{eff} \tau ^{\beta^{eff}(\Delta z,\Delta T)}
\end{eqnarray}
The value of $\beta^{eff}$($\Delta$z,$\Delta$T) essentially depends on the temperature interval $\Delta$T, used in the fit. It is close to the value of some true critical exponent if the fitting temperature interval $\Delta$T is outside the crossover regions. In Fig.15 we show the local effective critical exponent $\beta^{eff}$($\Delta$z,$\Delta$T) as a function of the distance from the pore center for 3 different temperature intervals $\Delta$T used for the fit. $\beta^{eff}$($\Delta$z,$\Delta$T) from averaging over the whole studied temperature range (see circles in Fig.15) shows a pronounced crossover from about $\beta$ = 0.38 in the pore center about 1.1 in the first surface layer. Moreover,  $\beta^{eff}$($\Delta$z,$\Delta$T) shows oscillations, which correlate with the density oscillations, observed in the liquid phase at low temperatures. In particular,  $\beta^{eff}$($\Delta$z,$\Delta$T) achieves the first local maximum of about 1.1 at the minimum of the fluid-wall potential. If the fitting temperature interval is reduced to 1.01 $\leq$ T $\le$ 1.12, the oscillations of $\beta^{eff}(\Delta z,\Delta T)$ become less pronounced and the effective exponent varies between 1 at the minimum of the fluid-wall potential to about 0.5 in the pore center. In the narrow high temperature interval 1.08 $\leq$ T $\le$ 1.12 the spatial crossover of $\beta^{eff}$($\Delta$z,$\Delta$T) is rather smooth, however, in the first surface layer $\beta^{eff}$($\Delta$z,$\Delta$T) is still higher (about 0.9) than in subsequent layers (see Fig.15). So, the spatial crossover of $\beta^{eff}$($\Delta$z,$\Delta$T) evidences that the localization of molecules near the wall due to the fluid-wall potential distorts the surface critical behavior toward higher values of $\beta_1$. A similar behavior was observed for the surface layer of water near a hydrophobic wall \cite{BGO2004a}. Away from the surface the density oscillations disappear at high temperatures and $\beta_1$ approaches the value $\beta_1$ = 0.82 of the ordinary surface transition of the Ising model.
\subsubsection{Density profiles of supercritical fluids} 
Density profiles of the LJ fluid confined in the slitlike pore were calculated for various average densities  along the bulk critical isotherm T$_c$ = 1.1876. At low densities the attractive fluid-wall potential caused adsorption of molecules near the surface and the profile in the pore interior is concave down (Fig.16,a). With increasing density the effect of density depletion near a weakly attractive substrate develops and provides density profiles which are convex up (Fig.16,b). The largest density gradient at T = T$_c$ roughly corresponds to the bulk critical density $\rho_c$ $\sim$ 0.3 $\sigma^3$. Further increase of the average fluid density in the pore results in density oscillations due to the packing effect (Fig.16,c). In the very dense supercritical fluid the density oscillations extend through the whole pore (Fig.16,d).
\begin{figure}
\includegraphics[width=7cm]{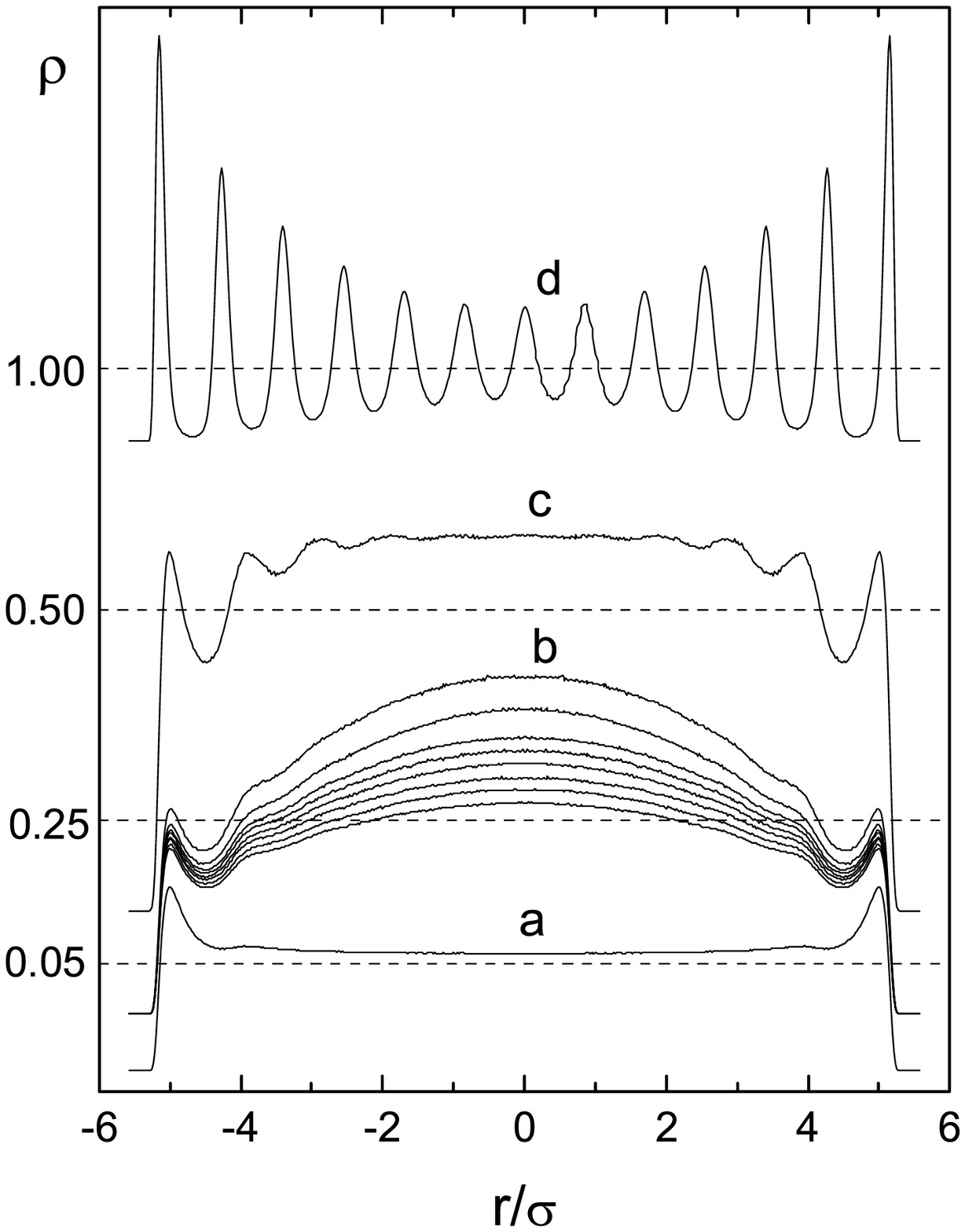}
\caption{Density profiles of LJ fluid confined in slitlike pore with H = 12 $\sigma$ along the bulk critical isotherm T$_c$ = 1.1876 at various average pore densities: $\rho$ = 0.05 $\sigma^3$ (a), 0.20 $\sigma^3$ $\leq\rho\leq$ 0.30 $\sigma^3$ (b), $\rho$ = 0.50 $\sigma^3$ (c) and $\rho$ = 1.00 $\sigma^3$ (d). The profiles b, c and d are shifted vertically. The averaged densities are shown by dashed lines. }
\end{figure} 
\begin{figure}
\includegraphics[width=7cm]{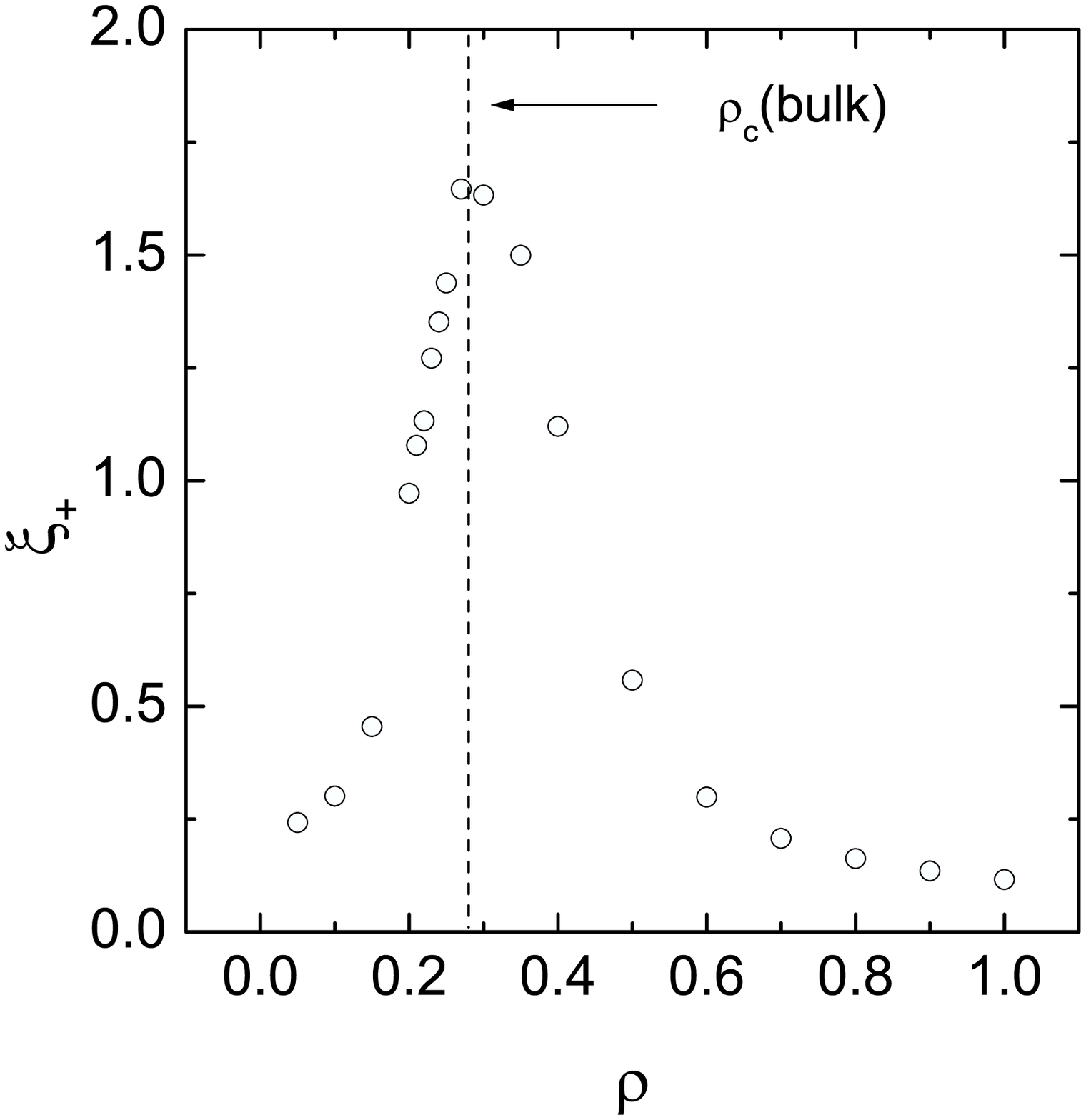}
\caption{Correlation length obtained from the fits of the Eq.(9) to the density profiles along the bulk  critical isotherm T$_c$ = 1.1876 (Fig.16).}
\end{figure}  
\par The density profiles at the bulk critical temperature from Fig.16 were fitted by Eq.(9), using $\rho_{bulk}$ and $\xi_+$ instead of $\Delta\rho_{bulk}$ and $\xi_-$, respectively. The parameter $\lambda$ was found close to zero, when the average pore density was below 0.15 $\sigma^3$ or above 0.5 $\sigma^3$. In the intermediate density range $\lambda$ passes trough a maximum of $\lambda$ $\approx$ 0.22 $\sigma$, when the pore average density is equal to 0.27 $\sigma^3$. Note, that a similarly small value of $\lambda$ was obtained from the master curve of the order parameter profiles (see Fig.9). The obtained values of the supercritical correlation length $\xi_+$ are shown in Fig.17. The maximum of $\xi_+$ is observed, when the average pore density is about 0.29 $\sigma^3$. When the correction factor, which accounts for the accessible volume, is applied to the average pore density (see section III.2), this value becomes practically equal to the bulk critical density $\rho_c\approx$ 0.32 $\sigma^3$. Note also, that the corresponding fitting parameter value $\rho_{bulk}\approx$ 0.42 $\sigma^3$ in Eq.(9) noticeably exceeds the critical density of the bulk LJ fluid . 
\begin{figure}
\includegraphics[width=8cm]{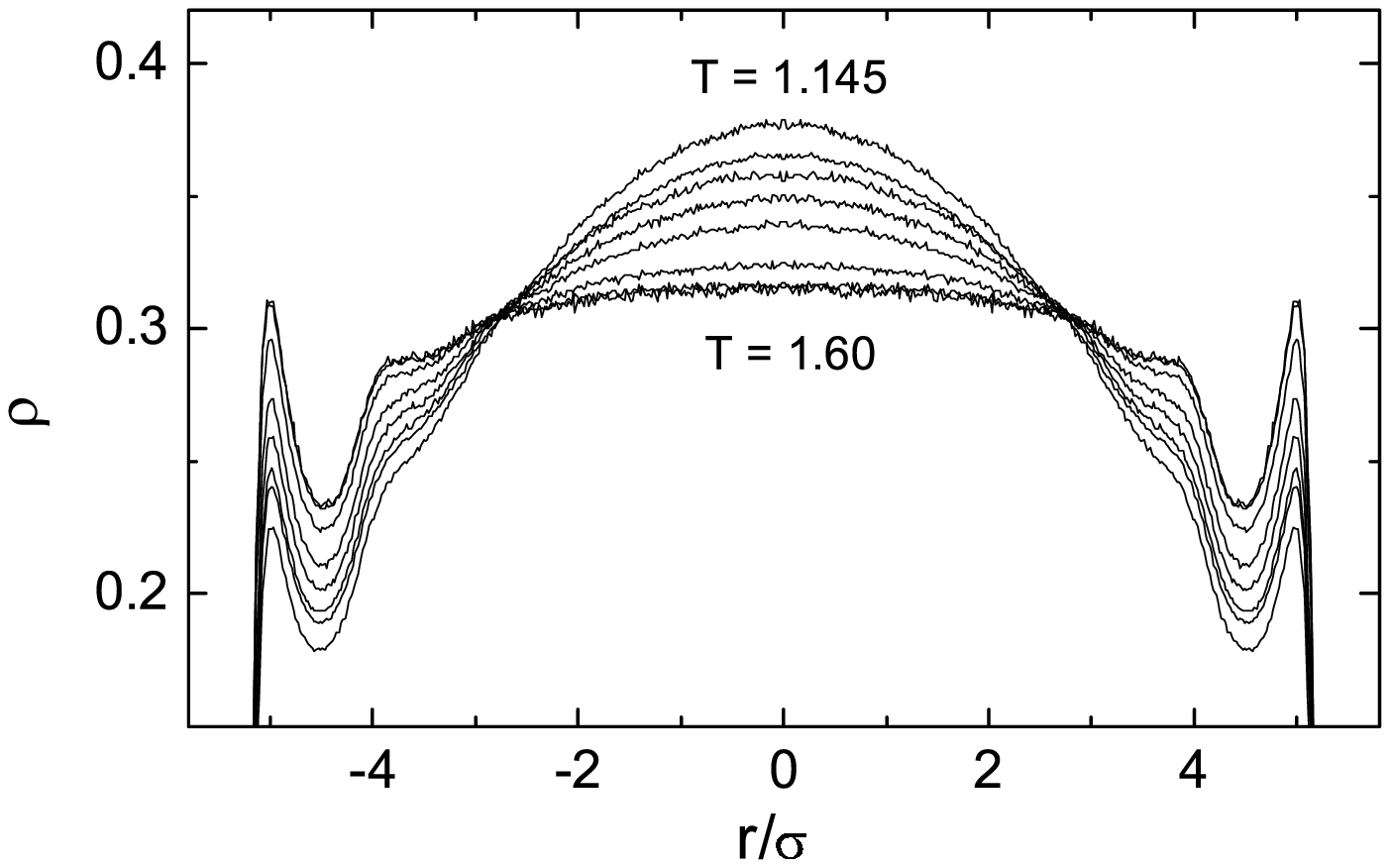}
\caption{Density profiles of LJ fluid confined in slitlike pore with H = 12 $\sigma$ along the pore critical isochore $\rho_c$ = 0.255 $\sigma^3$ at the temperatures T = 1.145, 1.1876, 1.20, 1.25, 1.30, 1.40 1.50 and 1.60.}
\end{figure}
\par 
Also, the temperature evolution of the density profile at supercritical temperatures was studied along the pore critical isochore $\rho_c$ = 0.255 $\sigma^3$ (Fig.18). All profiles cross in a point, where the density is about the bulk critical density. The density gradient from the pore surface to the interior decreases with increasing temperature and achieves a maximum at about the pore critical temperature T = 1.145. Accordingly, the correlation length, obtained from fits of Eq.(9) to the density profiles, also achieves its maximum $\xi_+$ = 1.87 $\sigma$ at this temperature. The parameter $\lambda$ varies in the range 0.21 $\sigma$ $\leq$ $\lambda$ $\leq$ 0.27 $\sigma$. So, we may conclude, that the value of about 0.2 $\sigma$ for  $\lambda$ determines the effective fluid boundary, when the density oscillations and the details of the fluid-wall potential are weak or negligible. This is the case for the critical isotherm in a wide density interval around the critical value and for the critical isochore. Besides, neglecting oscillations of the order parameter profile also gives a similar value of $\lambda$ at the coexistence curve (Fig.9). 
\par The temperature dependence of the average and local pore density diameters and of the local density along the pore critical isochore is shown in Fig.19. The average density diameter as well as the diameters in the surface layer and in the pore interior are continuously crossing over to the average and local densities at supercritical temperatures. The pronounced cusp in Fig.19 indicates, that the anomaly of the pore diameter (fit 6 in Table III) originates from the behavior of the fluid in the pore interior.   
\section{Discussion}
Computer simulations of a LJ fluid and water \cite{BGO2004a} evidence a highly universal surface critical behavior of fluids near weakly attractive surfaces. The long-range fluid-wall interaction suppresses a drying transition and therefore the order parameter could be studied in a wide temperature range. The temperature dependence of the local densities of the coexisting liquid and vapor phases near the surface could be described by Eq.(1). Near the surface this equation is valid in the whole temperature range, where the liquid phase exists: from supercooled temperatures to close proximity of the liquid-vapor critical point ($\tau \geq$ 0.05). This surface critical behavior intrudes into the bulk fluid on distances determined by the bulk correlation length $\xi_-$. This means, that at any distance from the surface the local order parameter crosses over from bulk-like critical behavior with exponent $\beta$ = 0.326 to surface critical behavior with exponent $\beta_1$ $\cong$ 0.8. The temperature of this crossover increases with increasing distance to the surface. The width of the fluid layer which follows the surface critical behavior described by Eq.(1) is about 2$\xi_-$ and diverges when approaching the critical temperature as $\tau^{-\nu}$.
\par The estimated value of the surface critical exponent is in good agreement with the value $\beta_1$ = 0.82 obtained for Ising magnets at zero surface field (h$_1$ = 0) \cite{beta1}. The largest deviations from the Ising behavior are observed in the first surface layer, where the fluid-wall potential causes a strong localization of the molecules both in the liquid and vapor phases, which does not disappear even at supercritical temperatures (Fig.18). As a result, the surface critical exponent, estimated for the first surface layer is about 0.9 for water \cite{BGO2004a} and 0.9 - 1.0 for the LJ fluid (Figs.12 and 15). Obviously, this deviation reflects the trend toward mean-field behavior with $\beta_1$ = 1 due to the localization of the molecules near the minimum of the fluid-wall potential. Besides, the effective value of the local exponent $\beta^{eff}(\Delta z,\Delta T)$, determined in a wide temperature range, varies similarly to the density oscillations in the liquid phase due to the packing effect (Fig.15).
\begin{figure}   
\includegraphics[width=8cm]{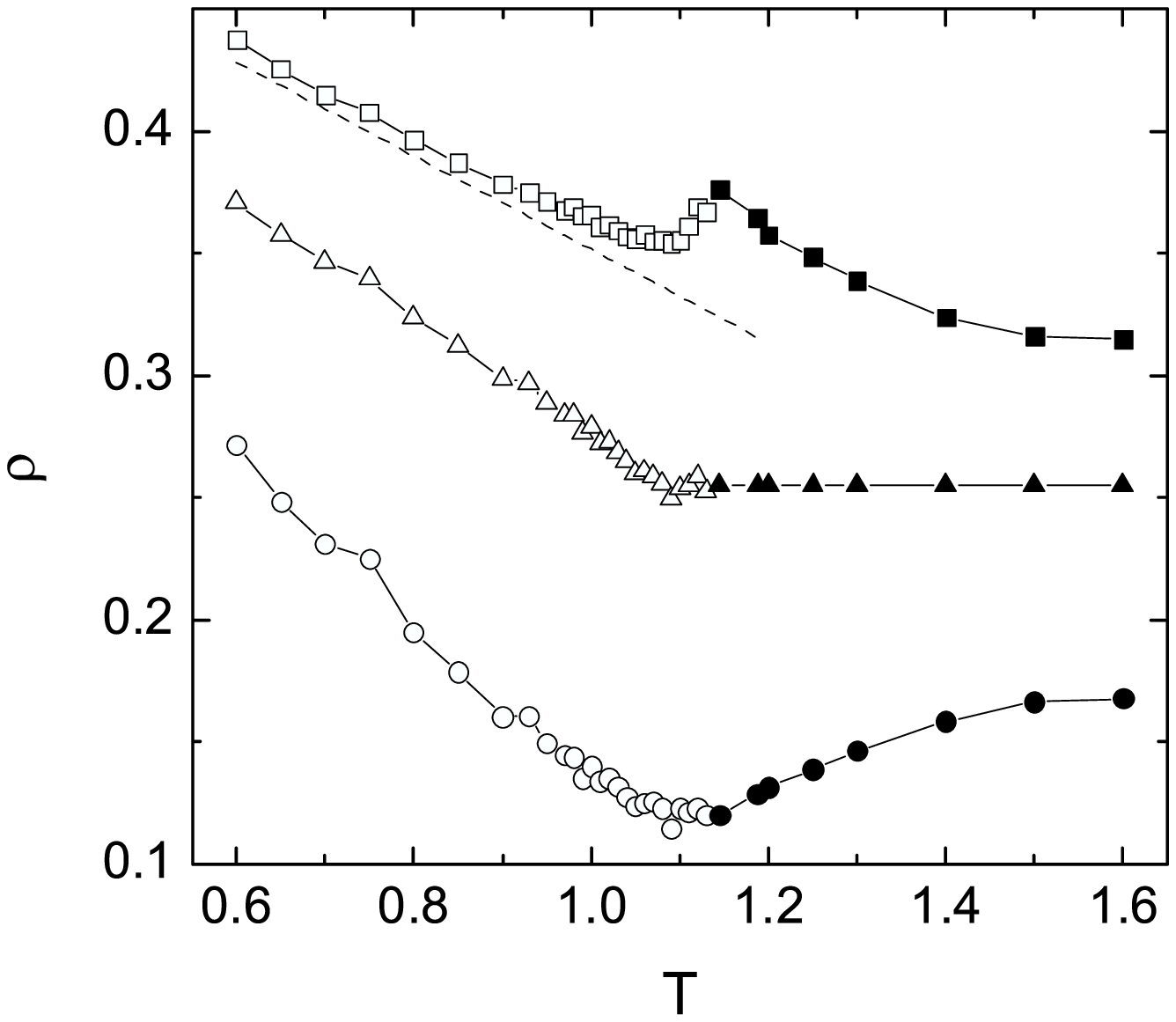}
\caption{Temperature dependence of the average pore density diameter (open triangles) and of the local density diameters $\rho_d$ in the surface layer (open circles) and near the pore center (open squares). The corresponding average and local densities at supercritical temperatures along the pore critical isochore $\rho$ = 0.255 $\sigma^3$ are shown by the respective solid symbols. The behavior of the diameter for the bulk LJ fluid is shown by the dashed line.
}
\end{figure}  
\par It may be assumed, that the difference $\Delta\rho$ between the densities of the coexisting phases near the surface should follow the behavior $\sim$ $\tau^{\beta_1}$ with $\beta_1$ = 0.82 also near strongly attractive surfaces below the wetting temperature. The contribution $\sim$ $\tau^{\beta_1}$ being present in both coexisting phases below the wetting temperature, as given in Eq.(1). Its disappearance in the liquid phase above the wetting temperature, when a liquid layer appears between the vapor and the solid surface, seems to be unlikely. So, the liquid density near the surface (both in the liquid phase and in the wetting layer) should include the positive contribution $\sim$ $\tau^{\beta_1}$. This means, that Eq.(1) could be valid also above the wetting temperature for the wetting phase. The absence of a vapor phase near the surface does not allow to separate symmetric and asymmetric contributions in Eq.(1), and so the extraction of the contribution $\sim$ $\tau^{\beta_1}$ is complicated by the presence of a linear regular term. In order to distinguish two contributions with close values of the exponents (0.8 and 1 in this case), a close approach to the critical point is needed. The available simulation methods may provide the high accuracy, necessary to solve this problem \cite{FisherLut}.
\par  
A fluid near an attractive wall in general should be mapped onto the $\textit{normal}$ transition universality class of the Ising model. The $\textit{normal}$ transition assumes that a non-zero surface field h$_1$ acts on the spins in the surface layer. If the surface field is absent (h$_1$ = 0), the Ising magnets show a surface critical behavior of the universality class of $\textit{ordinary}$ transitions. In the case of fluids, some strength of the fluid-wall interaction potential could provide equal attraction of the liquid and the coexisting vapor phase to the surface in the vicinity of the critical point. At this unique fluid-wall interaction neither an increase nor a depletion of the density is observed at the critical temperature \cite{Evans}. At any other fluid-wall interaction a preferential adsorption of one component (molecules or voids) occurs. The latter case corresponds to h$_1 \neq$ 0, i.e. to the $\textit{normal}$ transition in Ising magnets.
\par  In the case of the $\textit{normal}$ transition the magnetization in the surface layer m$_1$ in linear-response approximation contains regular and singular contributions above the critical temperature \cite{Diehlnorm1,Diehlnorm2}:
\begin{eqnarray}
m^+_1 = m_{1C} + A_1 \tau + ... + A_{2-\alpha}\tau^{2-\alpha}.
\end{eqnarray}
Below the critical temperature these contributions appear in the magnetizations of both coexisting phases. Due to the symmetry of the bulk Ising magnet the two phases contain the equal contributions (with same values and signs). However, at subcritical temperatures the surface magnetization should also include the term $\sim$ $\tau^{\beta_1}$, which describes the temperature dependence of the magnetization at h$_1$ = 0 (this term was overlooked in Refs.\cite{Diehlnorm1,Diehlnorm2}). This contribution accounts for the missing-neighbor effect and, therefore, has opposite signs in the two coexisting phases with magnetizations m$_1^{I}$ and m$_1^{II}$. So, the surface magnetization in the two phases along the coexisting curve should have the following temperature dependence:
\begin{eqnarray}
m_1^{I} = B_1\tau^{\beta_1} + m_{1C} + A_1\tau + ...+A_{2-\alpha}\tau^{2-\alpha}, \\ 
m_1^{II} = -B_1\tau^{\beta_1} + m_{1C} + A_1\tau + ...+A_{2-\alpha}\tau^{2-\alpha}.
\end{eqnarray}
In the presence of a non-zero surface field h$_1$ the two phases of the Ising magnet are no more symmetrical. Moreover, the magnetization near the surface does not vanish at the critical point. The order parameter of a phase transition is a measure for the dissimilarity of the two coexisting phases, which should vanish at the critical point. In the case of $\textit{ordinary}$ and $\textit{special}$ transitions \cite{Binderrev1} both bulk and surface magnetizations are equal to zero at the bulk critical temperature and therefore the magnetization is indeed the proper order parameter for these transitions. In the case of an $\textit{extraordinary}$ transition, the surface magnetization vanishes at the surface critical point, while at the bulk critical point it remains nonzero. Strictly speaking, the bulk magnetization serves as an order parameter for the bulk phase transition, while the surface magnetization is the proper order parameter for the surface phase transition, the critical point of which is located at higher temperature than T$_c$. In the case of the $\textit{normal}$ transition there is a single (bulk) phase transition and the order parameter should vanish at the bulk critical temperature. This means, that the magnetization could not be used as an order parameter for the $\textit{normal}$ transition, as it was done in the available theory of the $\textit{normal}$ transition \cite{Diehlnorm1,Diehlnorm2}.
\par
We propose to use as the order parameter of the phase transition in Ising magnet in all cases:
\begin{eqnarray}
\Delta m(z,\tau) = (m^{I}(z,\tau) - m^{II}(z,\tau))/2.
\end{eqnarray}
This order parameter is the deviation of the magnetization in each coexisting phase from the average magnetization at the same temperature:
\begin{eqnarray}
m_d(z,\tau) = (m^{I}(z,\tau) + m^{II}(z,\tau))/2.
\end{eqnarray}
In particular, the generalised order parameter $\Delta$m(z,$\tau$) and the diameter m$_d$(z,$\tau$) have the following temperature dependence in the surface layer :
\begin{eqnarray}
\Delta m_1(\tau) = B_1\tau^{\beta_1} ,   
\end{eqnarray} 
\begin{eqnarray}                            
m_{1,d}(\tau) = m_{1C} + A_1\tau + ...+A_{2-\alpha}\tau^{2-\alpha}.
\end{eqnarray} 
\par
In Ising magnets the order parameter $\Delta$m(z) is zero everywhere at T = T$_c$ at any surface field, excluding the case of the $\textit{extraordinary}$ transition, where $\Delta$m(z) vanishes at the surface critical point. Note, that the proposed generalisation (Eq.(15)) is equivalent to the standard definition of the order parameter, i.e. the magnetization, in the bulk case as well as in the cases of $\textit{ordinary}$, $\textit{extraordinary}$ and $\textit{special}$ transitions, because of the symmetry of the coexisting phases (m$^{I}$ = - m$^{II}$).
\par 
The average magnetization defined by Eq.(16) (i.e. diameter of the coexistence curve) is equal to zero in bulk Ising magnets and also near the surface with h$_1$ = 0. In the case of a $\textit{normal}$ transition m$_d$(z) reflects the response of the system to the surface field and describes the preferential adsorption of one kind of spins. The profile m$_d$(z) remains nonzero at T $\Rightarrow$ T$_c$ and continuously crosses over to the magnetization profiles in the supercritical region, showing critical adsorption. 
\par
We conclude, that in the absence of surface transitions the surface critical behavior of Ising magnets  belongs to the universality class of $\textit{normal}$ transition, described by Eqs.(13,14). The $\textit{ordinary}$ transition, which occurs at h$_1$ = 0 can be considered as a particular case of the $\textit{normal}$ transition, when  m$_d$(z) = 0. The proposed definitions of the order parameter and diameter (Eqs.(15,16)) allow to map the surface critical behavior of fluids onto the $\textit{normal}$ transition in Ising magnets, preserving isomorphism of the critical behaviors of bulk fluids and magnets. 

\section{Acknowledgments}
\par 
This work was supported by DFG Schwerpunktprogramm 1155. The authors thank K.Binder for advice and discussion of the $\textit{normal}$ transition in Ising magnets and also H.W.Diehl for stimulating discussions. 
\bibliography{len36}
\end{document}